\documentclass{article}

\usepackage{arxiv}

\usepackage[utf8]{inputenc} % allow utf-8 input
\usepackage[T1]{fontenc}    % use 8-bit T1 fonts
\usepackage{hyperref}       % hyperlinks
\usepackage{url}            % simple URL typesetting
\usepackage{booktabs}       % professional-quality tables
\usepackage{amsfonts}       % blackboard math symbols
\usepackage{nicefrac}       % compact symbols for 1/2, etc.
\usepackage{microtype}      % microtypography
\usepackage{lipsum}
\usepackage{graphicx}
\usepackage{chngcntr}
\graphicspath{ {./images/} }
\usepackage{natbib}
\usepackage{float}
\usepackage{longtable}
\usepackage{caption}
\usepackage{setspace}
\usepackage{subcaption}
\usepackage{placeins}
\setcitestyle{authoryear,open={(},close={)}}
\title{Forecasting AI Progress: Evidence from a Survey of Machine Learning Researchers}

\author
{Baobao Zhang,$^{1,2\ast}$ Noemi Dreksler,$^{1}$ Markus Anderljung,$^{1}$  Lauren Kahn,$^{3}$ \\ \textbf{Charlie Giattino,$^{4}$ Allan Dafoe,$^{1,5}$ Michael C. Horowitz$^{6}$} \\
\\
\normalsize{$^{1}$Centre for the Governance of AI}
\normalsize{$^{2}$ Maxwell School of Citizenship and Public Affairs, Syracuse University}\\
\normalsize{$^{3}$Council on Foreign Relations}
\normalsize{$^{4}$Our World In Data, University of Oxford}
\normalsize{$^{5}$DeepMind}\\
\normalsize{$^{6}$Perry World House and Department of Political Science, University of Pennsylvania}
\\
\normalsize{$^\ast$Correspondence to: baobaozhangresearch@gmail.com}
}

\begin{document}

\maketitle


\begin{abstract}
Advances in artificial intelligence (AI) are shaping modern life, from transportation, health care, science, finance, to the military. Forecasts of AI development could help improve policy- and decision-making. We report the results from a large survey of AI and machine learning (ML) researchers on their beliefs about progress in AI. The survey, fielded in late 2019, elicited forecasts for near-term AI development milestones and high- or human-level machine intelligence, defined as when machines are able to accomplish every or almost every task humans are able to do currently. As part of this study, we recontacted respondents from a highly-cited study by \cite{GraceEtAl} in which AI/ML researchers gave forecasts about high-level machine intelligence and near-term milestones in AI development. Results from our 2019 survey show that, in aggregate, AI/ML researchers surveyed placed a 50\% likelihood of human-level machine intelligence being achieved by 2060. The results show researchers newly contacted in 2019 expressed similar beliefs about the progress of advanced AI as respondents in the \cite{GraceEtAl} survey. For the recontacted participants from the \cite{GraceEtAl} study, the aggregate forecast for a 50\% likelihood of high-level machine intelligence shifted from 2062 to 2076, although this change is not statistically significant likely due to the small size of our panel sample. Forecasts of several near-term AI milestones have reduced in time, suggesting more optimism about AI progress. Finally, AI/ML researchers also exhibited significant optimism about how human-level machine intelligence will impact society.
\end{abstract}

\renewcommand{\figurename}{Fig.}

% In setting up this template for *Science* papers, we've used both
% the \section* command and the \subsection command for topical
% divisions.  Which you use will of course depend on the type of paper
% you're writing.  Review Articles tend to have displayed headings, for
% which \section* is more appropriate; Research Articles, when they have
% formal topical divisions at all, tend to signal them with bold text
% that runs into the paragraph, for which \subsection is the right
% choice.  Either way, use the asterisk (*) modifier, as shown, to

\section{Introduction}

Advances in artificial intelligence (AI) are shaping modern life, from transportation, health care, and science, to finance and the military. Understanding the trajectory of AI progress and monitoring its development and impact are critical tasks needed for proper governance of the technology \citep{Gruetzemacher2021researchagenda, whittlestone2021}. But forecasting technical progress in an area like AI/ML, just as in other areas, is extremely difficult, especially over longer time periods \citep{SuperforecastingTetlock,zhang2021ai, cremer_democratising_2021}. 

%The two previous AI winters, when disappointment with AI progress led to substantial reductions in funding for technological advances, highlights the way gaps can open up between expectations and reality. 

%Expert forecasts represent a fruitful avenue for understanding potential AI progress because they give us the ability to see similarities and differences in expectations across a subset of the AI/ML expert community, a group that is well positioned to grasp what types of breakthroughs are more or less likely in the coming years. Nevertheless, they have to tempered by the limitations of expert forecasts and the inherent challenges of predicting uncertain future events before using them as a precise measure of when certain AI milestones may be achieved.

In this paper, we report forecasts from a sample of 296 researchers who presented at two important AI/ML conferences --- the International Conference on Machine Learning (ICML) and Conference on Neural Information Processing Systems (NeurIPS) --- and compare our results to a survey of a similar population conducted in 2016 \citep{GraceEtAl}. The AI/ML researchers surveyed made forecasts about a wide range of AI capabilities, including when machines would be able to collectively perform almost all economically relevant tasks --- a notion of ``human-level machine intelligence.'' In this paper, we use the term \textit{HLMI} as an abbreviation for both human- and high-level machine intelligence; however, we specify ``human-level'' or ``high-level'' when making comparisons between these definitions. In addition to forecasts about HLMI, the respondents also made predictions about near-term AI progress milestones, such as when AI will be able to write a bestselling novel or beat the best human players at a particular video game.

%Respondents were also asked how positive or negative they would expect the overall impact of this human-level machine intelligence to be for humanity, in the long run, were it to exist ($n = 373$).

The survey recontacted respondents from \cite{GraceEtAl}, allowing us to compare 49 individually matched responses across the two surveys, three years apart. To our knowledge, this is the first time that a set of respondents has been recontacted for an AI progress forecast, offering a unique opportunity to study how AI/ML researchers' forecasts and views about the technology change over time.

%Nevertheless, there are limitations to expert forecasts given the inherent challenges of predicting uncertain future events, as our study demonstrates. Even so, our survey results could be useful for understanding how AI/ML researchers, influential actors in shaping the technology, think about near-term and long-term developments in their field. Their mental model of how AI would advance in the future could influence their research priorities \citep{prunkl2020beyond}. Furthermore, our study allows us to contrast how AI/ML researchers perceive the future of AI compared with other major stakeholders, such as the public.

Three key results from our 2019 survey can help us better understand how AI/ML will shape the future and in what time frame. First, in aggregate, the AI/ML researchers newly contacted in 2019 placed a 50\% probability of human-level machine intelligence being achieved by 2060. There is no significant difference to the results in \cite{GraceEtAl} from the 2016 survey, in which respondents predicted a 50\% likelihood of high-level machine intelligence by 2058. This lack of difference shows stability in forecasts about high or human-level machine intelligence over time. However, for \textit{recontacted} participants from the \cite{GraceEtAl} survey, the aggregate forecast for 50\% probability of high-level machine intelligence shifted from 2062 to 2076.

%We also show that AI/ML researchers' forecasts are susceptible to framing effects. In our study, we randomly assigned respondents to two ways to input their forecasts. In the \textit{fixed probabilities} framing, respondents were asked to input the years in which the milestone would be achieved with 10\%, 50\%, and 90\% probabilities. In the \textit{fixed years} framing, respondents were asked to input the likelihood the milestone would be achieved by a fixed set of years. The strongest predictor of a researchers' timeline for human-level machine intelligence was whether they were randomly assigned to give their forecasts using the fixed probabilities framing. The aggregate results shifted from a 50\% chance of human-level machine intelligence in 50 years with the fixed-year framing to a 50\% chance in 30 years with the fixed-probability framing. 

%% BZ 29-01-2022: I am moving the information about HLMI impact to the appendix to reduce the word count.

Second, our 2019 survey respondents appeared optimistic about how advances in AI/ML will impact humanity. They predicted that HLMI will be net positive for humanity, with the expected value between ``on balance good'' and neutral. The median AI/ML researcher ascribed a probability of 20\% that the long-run impact of HLMI on humanity would be ``extremely good (e.g., rapid growth in human flourishing)'', 27\% that it would be ``on balance good'', 16\% that it would be ``more or less neutral'', and 10\% that it would be ``on balance bad''. The median respondent placed a 2\% probability on HLMI being an ``extremely bad (e.g., human extinction)''.  

%A notable 11\% of respondents believed that there was a 100\% chance that the impact of HLMI on humanity would be ``on balance good'' or ``extremely good''. Ten percent of respondents placed above 50\% probability on the impact of HLMI being ``on balance bad'' or ``extremely bad.'' These results point to the possibility that AI researchers are more optimistic than the US public about the future impact of HLMI. \citep{zhang2019artificial} found in a poll of the US public that when the same question was posed as a multiple choice question \footnote{The question and response options were the same in both surveys, however, \citep{zhang2019artificial} asked the public to choose between the five outcomes while our survey of AI/ML researchers elicited probabilities for each outcome}, 34\% thought HLMI would be on net bad (22\% "on balance bad" and 12\% ``extremely bad'') and 26\% thought it would be on net good (21\% ``on balance good'' and 5\% ``extremely good''). A Pew Research survey of the general public in 20 different countries found, similarly, that 53\% of respondents said AI had been a good thing thus far, and 33\% that it had been a bad thing \citep{johnson_tyson_2020}.

Third, the 2019 survey respondents predicted significant progress across multiple domains up to 2080, believing that many AI milestones will be achieved faster than AI/ML researchers believed in 2016. For example, in the 2019 survey, the respondents predicted that an AI will write a \textit{New York Times} bestselling novel by 2034 with 50\% probability, compared to 2059 in the 2016 survey, possibly reflecting transformer-based advances in natural language processing \citep{devlin-etal-2019-bert}. 2019 respondents also appear more optimistic about AI milestones such as assembling any LEGO set and composing a Top-40 hit song.

Overall, these results highlight the beliefs of AI/ML researchers about progress in AI over the next generation and the impact it will likely have on humanity. Understanding the attitudes of this population provides insight on how AI development could evolve, the future of the technology, and potential research priorities in the field. Furthermore, it could assist international organizations and governments as they attempt to develop AI regulatory strategies.

%Overall, the results illustrate the uncertainty that AI/ML researchers about the trajectory of AI/ML progress, as illustrated by the wide range of forecasts respondents gave about advanced AI and even nearer-term developments in AI. Compared with the US public (as reported in \citep{zhang2019artificial}), AI/ML researchers forecast advanced AI systems arrive decades later than what the public forecast.

\section{Literature Review}

Forecasting AI progress is challenging, but there is a growing body of literature on different approaches to generating accurate forecasts, and the variables that influence forecasting success \citep{SuperforecastingTetlock,identifyingcultivatingsuperforecasters,psychofintelligenceanalysis}. Quantitative analysis of both performance trends and algorithm-based systems have utility for AI forecasting \citep{AIImpacts_hardware_nodate,DualIndicators,EFFAIProgressMeasurement,AICompute}.

% One popular method is to quantitatively analyze trends in how inputs like computational resources and data translate into outputs like performance at certain tasks --- sometimes called the ``AI production function.'' There is also growing evidence that algorithm-based systems have utility for forecasting, given the ability to use machine learning to incorporate large data sets and find patterns that can help anticipate what will happen next \citep{IntelligentFashion,forecastingImpactAI}.

Since some aspects of AI progress are more difficult to quantify, another approach is to elicit the subjective beliefs of people, including those with topical expertise such as AI/ML researchers, to better harness the ``wisdom of crowds'' \citep{RevisitingGalton}. This can provide insight into complex problems, how experts envision the future will unfold, and attribute realistic odds to possible futures \citep{BetterCrystalBall,PotentialforCrowdsourcingIntelligence}. %Probabilistic forecasting seeks to answer the questions of not only what different futures might look like, but which are most likely.

%Attempts have been made to aggregate opinions and expectations of policymakers, superforecasters (individuals who are consistently able to make forecasts more accurately than others) \citep{SuperforecastingTetlock}, AI and ML researchers, and the general public, through prediction tournaments, projects and surveys \citep{WhatdoweknowAITimelines, EvidenceagainstCurrentMethods,TechCast}. However, it should be noted that when asked to make predictions three to five years in the future, the task proves so challenging that expert communities perform only marginally better than random guessing \citep{ExpertPoliticalTetlock,TrendingUpHorowitzTetlock}.

\textit{AI Impacts}, a research project focused on understanding the impact of HLMI, reported that as of 2018 there existed ``around 1,300 public predictions of when human-level AI will arrive, of varying levels of quality.'' These predictions ranged from individual statements to larger surveys \citep{aiimpacts_predictions_2015,AIImpacts_aitimelines_2015}. At the time of writing, we are aware of 19 surveys on HLMI timelines, the first being published as early as 1972, that asked respondents to predict when they expected HLMI or artificial general intelligence to be achieved\footnote{Definitions for these terms varied by survey.} \citep{TechUnemployment,EztioniThreattoHumanity}. There are two key takeaways from these existing surveys: 1) all aggregate predictions for human/high-level machine intelligence say it will be after 2030, and most fall between 2035 and 2050, and 2) since 2016, the forecasts have shifted farther into the future, with most aggregate estimates falling between 2040 and 2070.

\begin{figure}
    \centering
    \includegraphics{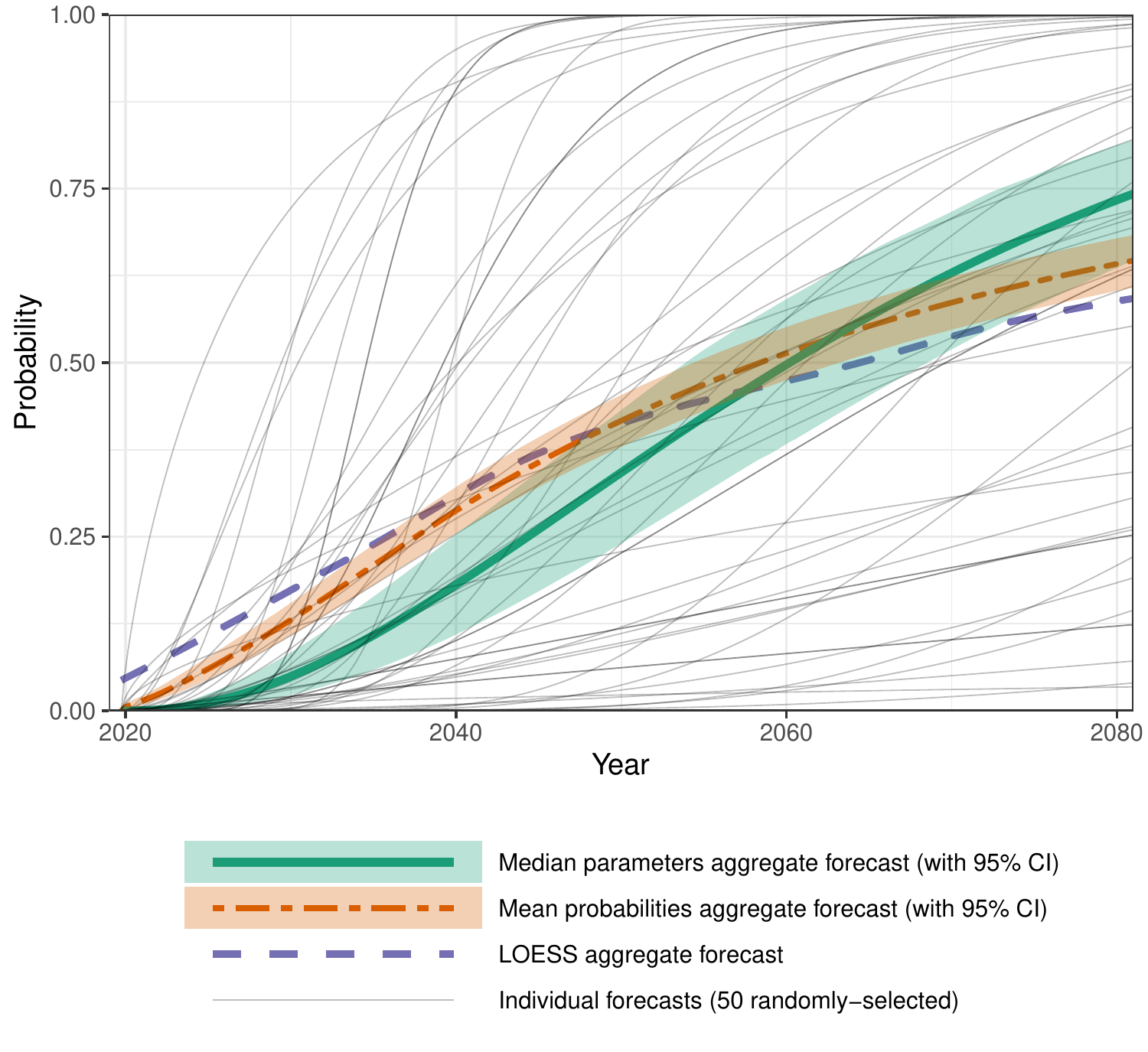}
    \caption{Forecast of human-level machine intelligence in the 2019 survey cross-sectional sample. The forecasts came from 289 individual respondents. We also show a random sample of the CDF forecasts of 50 individuals. The aggregated forecasts here imply a 50\% chance of human-level machine intelligence by 2060, with small variations depending on method.}
    \label{fig:cs_hlmi_fig_2019}
\end{figure}

\section{Materials and Methods}

\subsection{Sample}

\subsubsection{Cross-sectional Sample}

For our cross-sectional sample, we contacted all publishing authors (\textit{n} = 2652) at the 2018 meetings of two of the premier AI and machine learning (ML) conferences: the International Conference on Machine Learning (ICML; took place 10-15 July 2018) and the Annual Conference on Neural Information Processing Systems (NeurIPS; took place 2-8 December 2018). The survey was sent out and completed between 16 September and 13 October 2019. Due to a technical error, the surveys for part of our sample were missing some questions. To fix this problem, we sent out a corrected survey that was completed between 10-14 March 2020. We received a total of 524 responses (rate of ~20\%) which, after implementing our exclusion criteria,\footnote{Respondents were not required to answer every question, but we exclude those who did not answer the forecasting questions here (\textit{n} = 214). Other exclusion criteria were for responses that were not monotonically increasing (as a cumulative probability distribution should be; \textit{n} = 12) or contained unintelligible text rather than numeric values (\textit{n} = 2).} resulted in 296 responses included in the forecasting analyses here (198 from ICML and 108 from NeurIPS, including 10 who published at both). For the question on the impact of HLMI on the long run future of humanity, 373 individuals responded to the question across the cross-sectional and panel sample.

We describe the demographic characteristics of the respondents and non-respondents in Table \ref{tab:summarystat-cs}. A multiple regression that examines the association between demographic characteristics and response found that respondents have lower h-indexes (a measure of productivity and citation impact of researchers) and were more likely to work in academia compared with non-respondents (see Table \ref{tab:cs-response-regression-analysis}). Overall, however, we did not see evidence of concerning levels of response bias. Compared with other work of its kind, our 2019 survey has more respondents, a higher response rate, and more global coverage than other surveys of AI/ML researchers we reviewed. Separately from response bias, there are other aspects of the population of AI/ML researchers worth keeping in mind, such as gender imbalance (91\% of our respondents and 89\% of non-respondents were male, reflecting the low gender diversity of the field itself).

\subsubsection{Panel Sample}

Additionally, to compose our panel sample we recontacted all 352 respondents from \cite{GraceEtAl} to study changes in forecasts over time. We received a total of 84 responses (rate of ~24\%) which, after implementing exclusions, resulted in 49 responses that could be individually matched to responses from \cite{GraceEtAl} and that are included in the analyses here. The \cite{GraceEtAl} survey responses were collected between 3 May and 28 June 2016, giving roughly 3-3.5 years between responses for the panel sample. For tables and figures of AI milestones the difference used was 3.35 years (the difference between the midpoint of data collection for \cite{GraceEtAl} and the median time individuals completed our 2019 survey). 

Demographic characteristics of the respondents and non-respondents can be found in Table \ref{tab:summarystat-panel-sample}. The demographic data used to analyze whether the panel sample was representative of the 2016 survey respondents were collected for \cite{GraceEtAl}. Examining each demographic variable separately, there was a statistically significant difference in the percentage of respondents who received their undergraduate degrees in Asia (29\% among non-respondents and 14\% among respondents). A multiple regression that examined the association between demographic characteristics and response (reported in Table \ref{tab:panel-who-answered}) revealed that there was no statistically significant association between any of the demographic variables used and responding to the 2019 survey.  

\subsection{Pre-analysis Plan}

The pre-analysis plan for our 2019 survey can be found on the Open Science Framework: \url{https://osf.io/b239h/}. We largely followed our pre-analysis plan. We deviated from the plan by using a new aggregation method (the median parameters method) on our forecasting data. In the next subsection, we explain why we decided to use the median parameters method. As a robustness check, we also performed the analysis using the aggregation method described in our pre-analysis plan (the mean probabilities method). 

\subsection{Analysis of Forecasting Data}

While discrete forecasts (like the three probability or year values we elicited) can be useful, it is more informative to use these discrete values to construct a continuous probability distribution of an event happening over time (termed a cumulative distribution function or CDF). One major advantage of modeling forecasts as CDFs is that, unlike discrete forecasts, they are not susceptible to ``close-call counterfactuals'' in which a forecast outcome comes close to occurring by a certain date but does not \citep{Wallesten2016etal}.\footnote{For example, say a forecast question asks for the probability that Event X will happen by 7 July 2020. If Event X does not happen by 7 July but instead on 14 July, that is a close-call counterfactual. A forecast that predicted a relatively high probability for Event X happening by 7 July would receive a relatively bad score because the forecast outcome was that Event X did not happen by then. Yet it is clear to most people that the forecast was actually fairly close, especially if it was made, say, one year before. These situations are difficult to deal with using discrete forecasts but are not an issue with continuous forecasts/CDFs, because one has a probability value for every date (in this case) over a wide range.
} For the median parameters method, we constructed a CDF for each individual forecaster by fitting their three discrete values to a gamma CDF (defined by two parameters, shape and scale) using least squares optimization (the \textit{optim} function in R) \citep{GraceEtAl}. We then aggregated the individual forecasts into a group CDF by taking the median of the shape and scale parameters separately \citep{TidwellWallstenMoore2020,Tidwell2017}. We generated 95\% confidence intervals (CIs) for the group median CDF by bootstrapping (10,000 simulations) at the forecaster level.

Our median parameters method for group aggregation differs from the method used in \cite{GraceEtAl}, which took the mean of the cumulative probabilities at each year. We chose this new method because there is evidence it leads to more accurate forecasts \citep{Tidwell2017,TidwellWallstenMoore2020}, is computationally simpler, and is less sensitive to outliers and more resistant to compression to the center of a probability distribution \citep{Baron2014}. The current ``median of gamma parameters'' method and the \cite{GraceEtAl} method led to different forecasts as can be seen in Fig.  \ref{fig:cs_hlmi_fig_2019}. We also aggregated the forecasts using locally estimated scatterplot smoothing (LOESS), which does not require us to make assumptions about the functional form of the forecasts. We chose smoothing hyperparameters for the LOESS as to minimize a criterion based on the Akaike information criterion \citep{hurvich1998smoothing}. 

Building on the description of Fig. \ref{fig:cs_hlmi_fig_2019}, the median parameters method resulted in a steeper aggregate CDF than the locally estimated scatterplot smoothing (LOESS) method. Furthermore, the median parameters method also generated a steeper aggregate CDF than the method that takes the mean of the cumulative probabilities at each year (hence known as the \textit{mean probabilities method}) used in \cite{GraceEtAl}. The 50\% probability predicted year, however, remained remarkably close between the mean and median methodologies for the 2019 data.

It is worth noting that the median parameters method fits a gamma function to the data, which assumes that the probabilities sum at 100\% eventually. This is strictly speaking of course not true: respondents may believe that there is a chance that HLMI may never be achieved. It is likely this behavior of the gamma function is in part to blame for the steeper CDFs that the median parameter method obtains and should be kept in mind when interpreting the tails of the CDFs that lie far into the future.

\section{Results}

\subsection{Human-level Machine Intelligence Forecasts from 2019 Cross-sectional Sample}

We present the results below for human-level machine intelligence forecasts for 1) our main cross-sectional sample consisting of 289 authors that published at the 2018 NeurIPS and ICML conferences and 2) a panel sample made up of 49 respondents recontacted from a survey by \cite{GraceEtAl} conducted in 2016. The cross-sectional sample respondents were asked to forecast when \textit{human-level machine intelligence} would exist, defined as ``when machines are collectively able to perform almost all tasks ($>90\%$ of all tasks) that are economically relevant better than the median human paid to do that task.'' To be consistent with the prior survey phrasing, we asked the panel sample respondents about \textit{high-level machine intelligence}, as in 2016, which was defined as when ``unaided machines can accomplish every task better and more cheaply than human workers'' and were asked to assume that scientific activity would continue without major negative disruption.

This question was asked in either a \textit{fixed years} or \textit{fixed probabilities} framing. The \textit{fixed years} framing asked for their subjective probability that HLMI would exist in a certain number of years (for the cross-sectional sample: 10, 20, 50 years, panel sample: 10, 20, 40 years.\footnote{In the 2016 survey, we asked respondents to provide probabilities for 10, 20, and 40 years. For comparability, we kept the fixed years the same in the 2019 survey for the panel sample.}) The \textit{fixed probabilities} framing presented three probabilities (10\%, 50\%, and 90\%) and asked both cross-sectional and panel sample respondents in how many years they would expect human or high-level machine intelligence to exist with that probability.

For the cross-sectional sample, the aggregate forecast across both framings for when high-level machine intelligence would exist with 50\% probability was 2060, according to the median parameters aggregation method. As Fig. \ref{fig:cs_hlmi_fig_2019} shows, the aggregate forecast for the cross-sectional sample in 2019 varied somewhat depending on which aggregation methodology was used.\footnote{See the supplementary materials for more detail regarding our aggregation methods.} For the parametric aggregation methods in Fig. \ref{fig:cs_hlmi_fig_2019}, we fit each respondent's forecasts to cumulative distribution functions (CDFs) following the gamma distribution, before aggregating these CDFs.

% %\footnote{When adjusted for when the survey was taken within the year (between 16 September and 13 October 2019), the aggregate forecast lies around 2060.},

\subsection{Changes in Human/High-Level Machine Intelligence Forecasts: 2016-2019}

\begin{figure}[ht]
    \centering
    \includegraphics{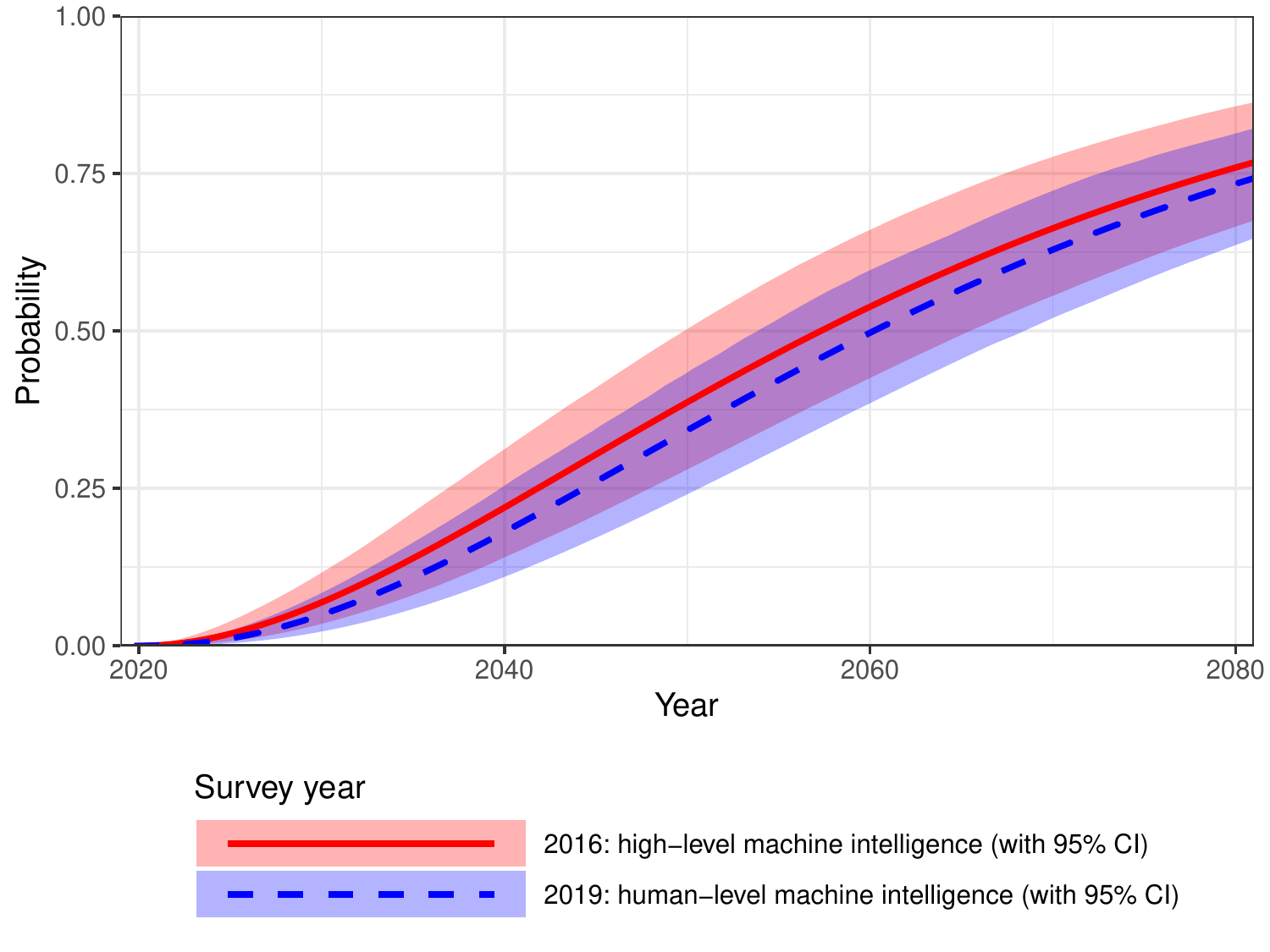}
    \caption{Comparing aggregate forecast CDFs in 2016 and 2019 using the median parameters method. The blue line represents all 2019 survey responses with the 2019 human-level machine intelligence definition and the red line represents all 2016 survey responses with the 2016 high-level machine intelligence definition.}
    \label{fig:2016v2019_hlmi}
\end{figure}

The results from 2019 showed consistency with the 2016 survey results, illustrating stability in how AI/ML experts think about the probability of HLMI. The aggregate 50\% probability of HLMI forecast shifted from 2058 to 2060 from the \cite{GraceEtAl} survey in 2016 to our survey in 2019.\footnote{The data from \cite{GraceEtAl} was re-analyzed using the median parameters method we use for the present data to make the results comparable.}  The CDFs of the aggregate forecasts by AI/ML researchers in 2016 and 2019 can be seen in Fig. \ref{fig:2016v2019_hlmi}. Randomization inference tests showed no significant differences in the aggregate median parameters CDFs between the AI/ML researchers sampled in 2016 and 2019.\footnote{The randomization tests used the Wasserstein metric and the Kolmogorov-Smirnov as test statistics of the difference between the two sets of forecasts. The test statistics were applied to the interval between the years 2019 and 2080. See the Supplementary Materials for the results of all significance tests of changes in the forecasts across the two surveys.}

To further evaluate the stability of high-level machine intelligence forecasts we evaluate the responses from 49 individuals who answered the same question framing in 2016 and 2019. As Fig. \ref{fig:hlmipanel2016v2019} in the Supplementary Materials shows, the panel sample's forecasts were pushed further into the future in the 2019 survey in comparison to the 2016 survey. The predicted 50\% probability of achieving high-level machine intelligence shifts from 2062 to 2076 for the panel sample.\footnote{Despite substantive changes between the two panels, due to the small size of the panel samples, randomization inference methods and additional tests suggest this difference is not statistically significant at the 5\%-level. The results are presented in the Supplementary Materials.}

%\subsection{What predicts HLMI timelines in our Cross-sectional Sample?}

Given the differences in human-level machine intelligence forecasts across respondents, we examined factors that could predict differences in forecasts. We ran three linear regression models (with robust standard errors) to predict the log value of the year when the probability of human-level machine intelligence is at 50\% in the cross-sectional sample (see Table \ref{predict-hlmi-table}). In the model using only demographic variable predictors, individuals who completed their undergraduate studies in Asia predicted significantly shorter timelines (49\% shorter) in comparison to researchers that completed their undergraduate degrees in North America, a finding that mirrors that of \cite{GraceEtAl}. In the model that includes both demographic predictors and features of the survey, only question framing significantly predicted timeline lengths. Like \cite{GraceEtAl}, we find that the fixed probabilities framing caused respondents to give shorter timelines (57\% shorter) than the fixed years framing.

\begin{figure}[H]
    \centering
    \includegraphics[width=\textwidth]{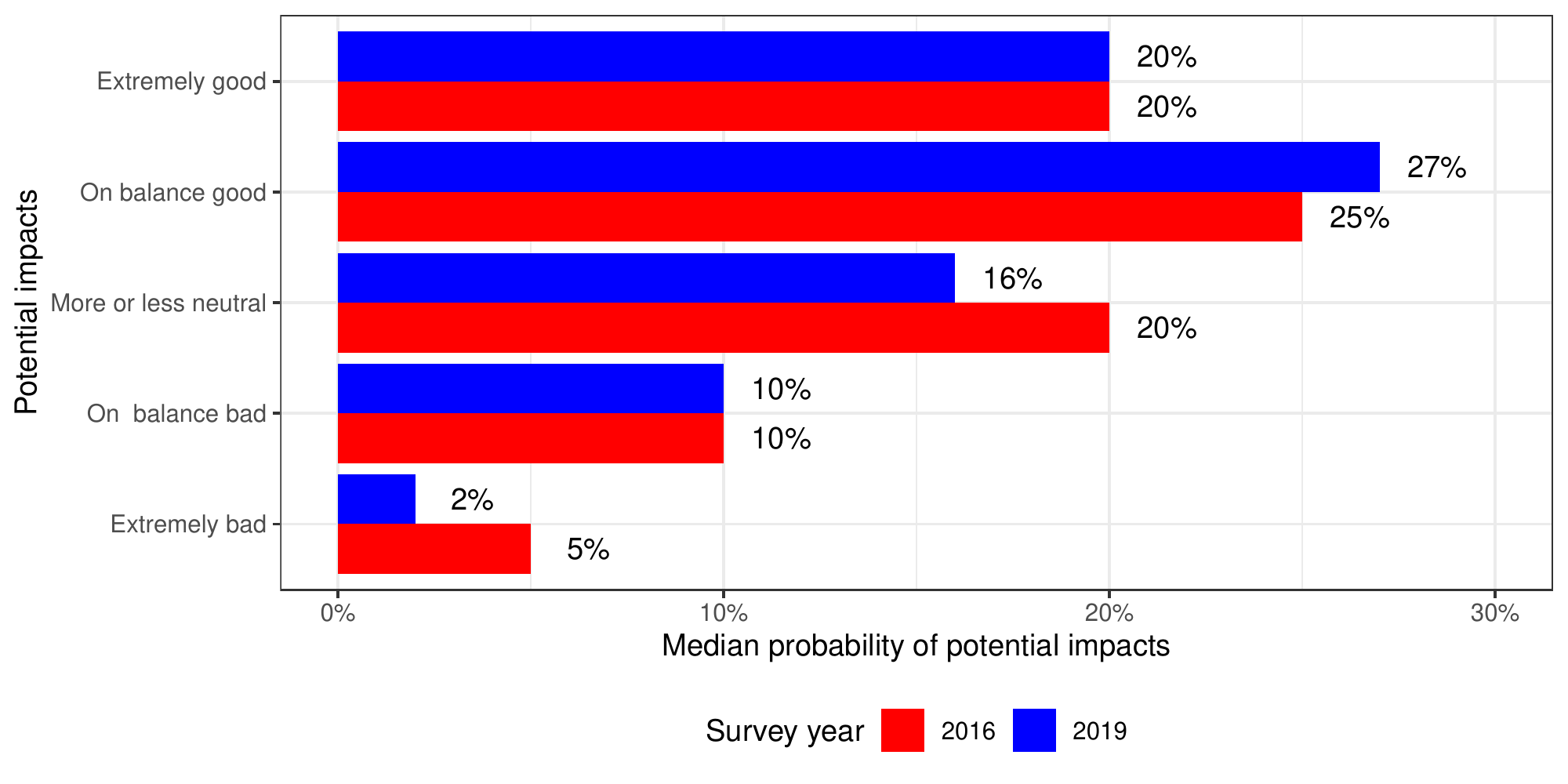}
    \caption{The probability that the median respondent placed on each outcome when asked about the potential impact of human/high-level machine intelligence (HLMI) on the long run future of humanity.}
    \label{fig:hlmi_impact_2016v2019medianprobabilities}
\end{figure}

\subsection{Impact of HLMI}

We asked respondents for the probabilities that they would place on the impact of HLMI on humanity, in the long run, being ``extremely good (e.g., rapid growth in human wellbeing),'' ``on balance good,'' ``more or less neutral,'' ``on balance bad,'' and ``extremely bad (e.g., human extinction).''  As with their forecasts for when HLMI will occur, the impact forecasts show consistency with the 2016 survey as can be seen in Fig. \ref{fig:hlmi_impact_2016v2019medianprobabilities}. 

Overall, AI/ML researchers have a net positive outlook on the potential impact of HLMI on humanity in the long run. The expected value of HLMI impact in 2019 is $0.59$ --- roughly midway between ``on balance good'' (1) and ``more or less neutral'' (0)\footnote{A robust linear regression showed no significant difference in expected value between the panel sample who was asked about high-level machine intelligence and the cross-sectional sample who was asked about human-level machine intelligence. We combined the data across the two samples in all analyses for this question.}. This is not significantly different to the $0.52$ expected value of the impact of HLMI that AI/ML researchers predicted in 2016 (\textit{p} = $0.08$).

% \begin{figure}
%     \centering
%     \includegraphics[width=\textwidth]{images/milestone_2016_2019_compare_non_sample.pdf}
%     \caption{Comparing 2016 and 2019 forecasts of AI development milestones, cross-sectional sample. The results are produced using the median parameters aggregation method. Each points represents the year at 50\% probability of achieving the milestone. The lower end of the confidence interval represents 25\% probability of achieving the milestone. The upper end of the confidence interval represents 75\% probability of achieving the milestone.}
%     \label{fig:milestone_compare_cs_sample}
% \end{figure}

\subsection{AI Progress Milestones}

AI progress milestones --- such as when AI will be able to write a best-selling novel or play a video game better than the best human players --- represent a concrete way to judge AI/ML progress. We asked each respondent to make forecasts about three AI progress milestones. For the cross-sectional sample, we included 18 milestones from \cite{GraceEtAl} and four additional ones.\footnote{We selected the four additional milestones after consulting with AI/ML researchers. We chose milestones that would have clear resolution criteria, are easily understandable, where AI researchers may be considered to have expertise, and where their resolution would say something important about AI progress.} For the panel sample, we asked respondents about the same milestones that they responded to in \cite{GraceEtAl}.

This yields an average of 36.5 forecasts per milestone. The 2016 survey asked respondents about when a milestone would be \textit{feasible} --- whether ``one of the best resourced labs could implement it in less than a year if they chose to'' --- whereas our 2019 survey asked respondents about whether the milestone would be \textit{achieved and made public} to make it easier to objectively determine whether a milestone forecast had resolved. Thus, when comparing the 2016 to the 2019 milestone forecasts, we can detect whether forecasts have moved closer to the present: if the 2019 \textit{achievability} forecast is sooner than the 2016 \textit{feasibility} forecast, we can be certain that researcher beliefs have changed. If the 2019 forecasts are further into the future, that may be a result of the respondents' beliefs about how much time will elapse between when a milestone can feasibly be met and when it is achieved.

\begin{figure}[H]
    \centering
    \includegraphics[width=0.9\textwidth]{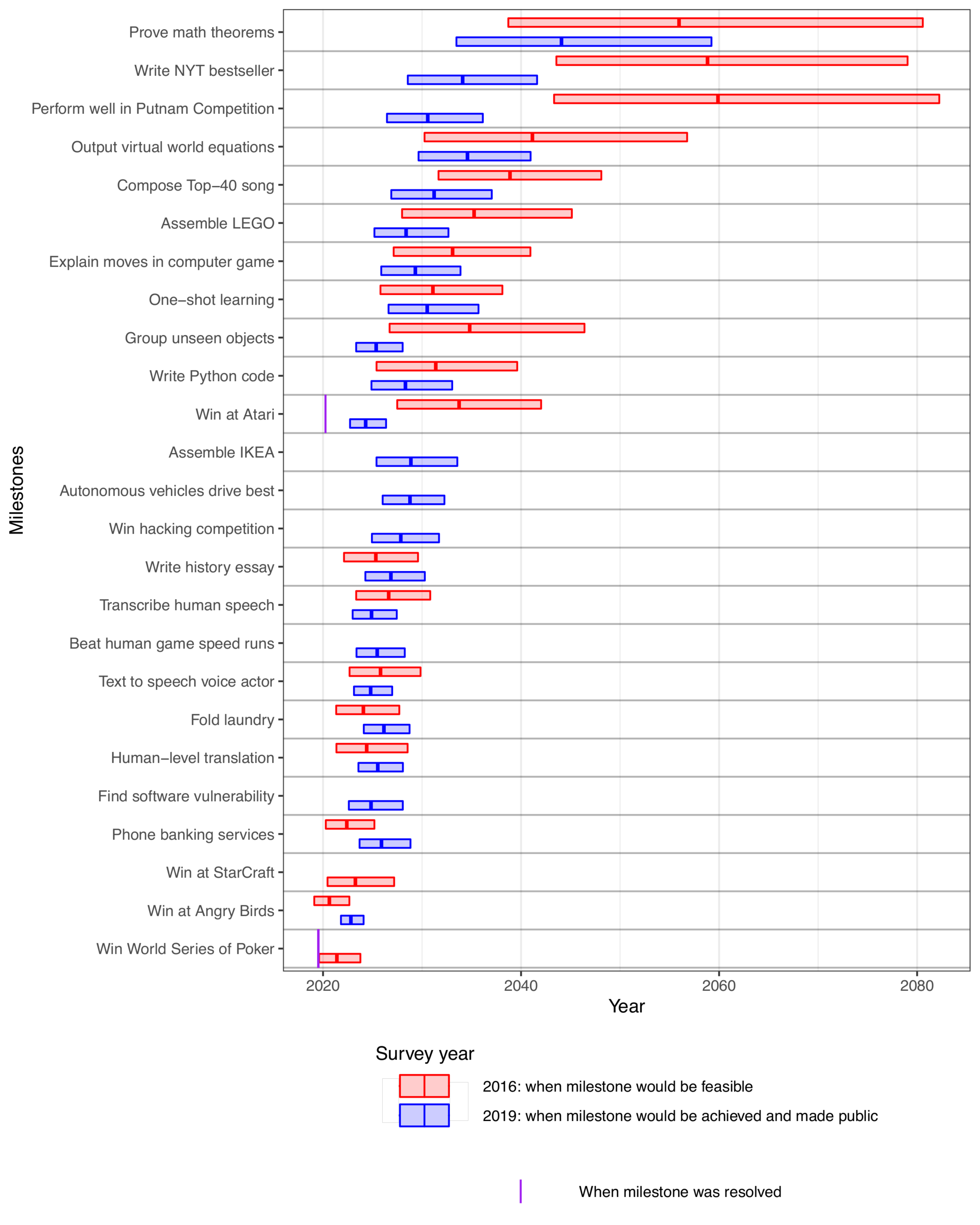}
    \caption{Comparing the 2016 and 2019 forecasts of AI progress milestones, cross-sectional sample. The results are produced using the median parameters aggregation method. The center line of each boxplot represents the year at 50\% probability of achieving the milestone. The lower end of the boxplot represents 25\% probability of achieving the milestone. The upper end of the boxplot represents 75\% probability of achieving the milestone.}
    \label{fig:milestone_compare_cs_sample}
\end{figure}

In Fig. \ref{fig:milestone_compare_cs_sample}, we compared the 2016 and 2019 forecasts from the cross-sectional samples. The results were produced using the same median parameters method used to aggregate the human/high-level machine intelligence forecasts.\footnote{The Supplementary Materials show the results using the mean probabilities aggregation method from \cite{GraceEtAl} as a robustness test.}. Overall, for 13 of the 18 milestone forecasts, researchers in the 2019 survey predicted that the milestones would occur earlier than researchers did in the 2016 survey. This includes milestones such AI proving math theorems published in leading math journals today (the forecast 50\% probability is 11.9 years earlier in 2019), writing a \textit{New York Times} bestseller (24.7 years earlier), or performing well in the Putnam Competition, a math competition for undergraduate students in North America (29.3 years earlier). This is especially noteworthy given the 2019 definition is harder to reach (i.e., achievability, rather than feasibility). 

Some of the forecasts are more comparable. In both years, AI outperforming the best human players at the Angry Birds AI Competition was predicted to be achieved the soonest (with 50\% probability by 2020.6 in the 2016 survey; with 50\% by 2022.8 in the 2019 survey).

%This is consistent with literature on forecasts, which has shown that the further out a prediction is, the less accurate and less precise they become. As \citep{SuperforecastingTetlock} explains,``the further out the forecaster tries to look, the more opportunity there is for chaos to flap its butterfly wings and and blow away expectations'' (pg. 14). It may also point to these more challenging milestones being seen as more achievable after the progress seen in the three years between the two surveys. Milestones which had generally closer timelines in 2016, were more likely to lengthen, than those which had very long timelines, which were more likely to become shorter and more aggressive.

Some of our formal statistical tests (in Table \ref{tab:npc-results}) point to differences between the 2016 and 2019 samples in the cross-sectional sample, with significance depending on the test statistic that is used (see Additional Key Results in the Supplementary Materials). We found no statistically significant difference in the confidence of forecasts (see Table \ref{tab:robust-cross-fp}). However, this does not necessarily mean the respondents have not become more confident in their views. It could be an artifact of our asking about achievability rather than feasibility in 2019: achievability ought to produce higher uncertainty forecasts than feasibility, as something is only achieved if it is feasible and an actor chooses to pursue the milestone. In the cross-sectional sample, there was no statistically significant correlation between the human-level machine intelligence and milestone forecasts (see Table \ref{tab:hlmi-milestone-regression} in the Supplementary Materials).

% We tested this by using the log year of achieving human-level machine intelligence with 50\% to predict the log year of achieving the milestone with 50\%. In one model, we used fixed effects to account for variations in milestones; in another model, we used random effects instead. In both models, we clustered the standard errors by respondent. In both models (Table \ref{tab:hlmi-milestone-regression}), we find no correlation between the log year of achieving human-level machine intelligence and the log year of achieving the milestones ($p$-value $= 0.695$ in the fixed effects model; $p$-value $= 0.785$ in the random effects model). 

\subsection{Milestone Resolutions}

Two AI milestones have been achieved since the 2016 respondents made their predictions: DeepMind's Agent57 \citep{agent57} beat the human Atari benchmark (published in March 2020) and Facebook and Carnegie Mellon's Pluribus \citep{pluribus} beat top human poker players at Texas hold 'em (published in July 2019). The 2016 respondents gave a 50\% chance that the \textit{Win at Atari} milestone would be met by late 2033 (aggregated using the median parameters method). Our 2019 respondents, on the other hand, making their forecasts just five to six months before the achievement would be publicly announced, produced a prediction of 50\% chance the milestone would be met by 2024 (aggregated using the median parameters method). For both surveys, the number of milestones resolved by January 1, 2022 is not statistically different than the number of milestones forecast to resolve (see Fig.  \ref{fig:forecasts-milestone-null-dist}). On average, the 2016 respondents forecast 3.92 milestones to resolve (95\% CI: 1--7) and two milestones from that survey resolved by January 1, 2022; similarly, the 2019 respondents forecast 1.65 milestones to resolve (95\% CI: 0--4) and one milestone from the survey resolved.

We did not ask respondents to predict the \textit{Win at Poker} milestone in 2019 as it had already been achieved at that point. In 2016, the respondents put a 50\% probability that the \textit{Win at Poker} milestone would be met by the middle of 2021, which is about two years after the actual achievement. As more of the milestones resolve, we intend to score researchers' forecasts and investigate what, if any, correlation exists between accurate predictions on individual milestones and forecasts of human-level machine intelligence. It is also likely that a number of the milestones are close to resolution or might be already be feasible (see additional discussion in the Supplementary Materials Additional Key Results section).

\section{Conclusion}

Forecasting AI progress remains both inherently difficult and critical for the future of humanity. Our study points to the stability of forecasts regarding human/high-level machine intelligence and nearer-term AI progress milestones across time for a subset of AI/ML researchers that published at NeurIPS and ICML. Results from our 2019 survey show that in aggregate AI/ML researchers surveyed place a 50\% chance of human-level machine intelligence being achieved by 2060. There is no significant difference to the results in \cite{GraceEtAl}. Additionally, recontacted respondents from the \cite{GraceEtAl} survey who responded to our 2019 survey did not significantly change their forecasts regarding high-level machine intelligence. Forecasts of near-term AI milestones did show a substantial movement towards the present, but the significance of the milestone differences between the 2016 and 2019 surveys varies.

There are limitations to this research, despite the findings and importance. First, the panel sample size ($49$ for the human-level machine intelligence forecasts; $41$ for the milestones forecasts) makes testing for within-respondent changes difficult. Second, our 2019 survey sought to describe AI progress milestones that were high-impact, easy for respondents to understand, and could plausibly be resolved in the real-world. Not all three criteria could be achieved at once. Finally, we point out that the forecasts made by our 2019 survey respondents are not necessarily accurate. For instance, forecasts regarding HLMI were subject to framing effects. Leading domain experts do not necessarily make the best forecasts, and predicting long term outcomes is difficult \citep{ExpertPoliticalTetlock, SuperforecastingTetlock}. Even so, the results allow us to understand what researchers directly building AI think about the future of the technology. Researchers' beliefs about AI progress timelines could potentially shape their research priorities.

Future research should explore a variety of avenues to advance our understanding of forecasting AI progress. First, it would be fruitful to follow larger samples over a longer period of time than three years to build a longitudinal picture of how HLMI forecasts are changing in individuals in response to AI progress. This would also allow us to get a better handle on whether the non-significant lengthening of high-level machine intelligence timelines we see in the panel sample points to an effect we should be taking note of. It could be particularly helpful to figure out whether notable changes in forecasts preempt significant advances to make it easier to predict impending breakthroughs. Second, future research should attempt to establish why the fixed years versus fixed probabilities framing effect is taking place and which method of eliciting forecasts from experts in surveys results in the most accurate and well-calibrated forecasts.

\section{Acknowledgments}

We want to thank Emmie Hine, Tegan McCaslin, Kwan Yee Ng, and Catherine Peng for their research assistance. For helpful feedback and input, we want to thank: Catherine Aiken, Carolyn Ashurst, Miles Brundage, Rosie Campbell, Alexis Carlier, Jeff Ding, Owain Evans, Ben Garfinkel, Katja Grace, Ross Gruetzemacher, Jade Leung, Alex Lintz, Max Negele, Toby Shevlane, Brian Tse, Eva Vivalt, Waqar Zaidi, Remco Zwetsloot, the OECD Future of Work Initiative, our colleagues at our respective institutions. We are also grateful for research support from the Center for Security and Emerging Technology at Georgetown University and the Berkeley Existential Risk Initiative. 

This study has received exemption from the University of Pennsylvania Institutional Review Board (Protocol \# 828933).

This research was supported by: the Ethics and Governance of AI Fund, the Open Philanthropy Project grant for ``Oxford University -- Research on the Global Politics of AI,'' the Minerva Research Initiative under Grant \#FA9550-18-1-0194, the CIFAR Azrieli Global Scholars Program, the OECD Future of Work Fellowship, and the Long-Term Future Fund.

%Here you should list the contents of your Supplementary Materials -- below is an example. 
%You should include a list of Supplementary figures, Tables, and any references that appear only in the SM. 
%Note that the reference numbering continues from the main text to the SM.
% In the example below, Refs. 4-10 were cited only in the SM.     
\appendix

\renewcommand\thefigure{S\arabic{figure}}   
\renewcommand\thetable{S\arabic{table}} 
\setcounter{figure}{0} 
\setcounter{table}{0}

% For your review copy (i.e., the file you initially send in for
% evaluation), you can use the {figure} environment and the
% \includegraphics command to stream your figures into the text, placing
% all figures at the end.  For the final, revised manuscript for
% acceptance and production, however, PostScript or other graphics
% should not be streamed into your compliled file.  Instead, set
% captions as simple paragraphs (with a \noindent tag), setting them
% off from the rest of the text with a \clearpage as shown  below, and
% submit figures as separate files according to the Art Department's
% instructions.

\clearpage

\section{Supplementary Materials}

\noindent Additional Key Results

\noindent Supplementary Text: Survey Text

\noindent Figs. S1 – S11

\noindent Tables S1 – S35

\newpage

\appendix

\section{Additional Key Results}

\subsection{What Predicts Human-Level Machine Intelligence Timelines in Our Cross-sectional Sample?}

What factors could drive differences in human-level machine intelligence timelines? We ran three linear regression models (with robust standard errors) to predict the log value of the year when the probability of human-level machine intelligence is at 50\% in the cross-sectional sample (see Table \ref{predict-hlmi-table}). Where more than 10\% of data was missing for a predictor we added an indicator variable that conditioned on this missing data. We emphasize that none of the significant results noted below would survive a Holm's correction for multiple comparisons. The reasons for this may be that the timelines simply are not well predicted by any of the variables in our regression model, or that the regression models are attempting to fit too many parameters, especially in comparison to the size of the sample. We present the results for the uncorrected regression models below but ask researchers to mention this caveat regarding multiple comparisons when citing the results.

\begin{table}[H]

\centering
\caption{Predicting log(year when probability of human-level machine intelligence is at 50\%) using demographic variables and framing experimental group; 2019 survey cross-sectional sample, 2019 human-level machine intelligence definition \label{predict-hlmi-table}}

\begin{tabular}{p{9cm}ccc}
\toprule
  & Model 1 & Model 2 & Model 3\\
\midrule
Intercept & 4.208*** & 2.876* & 3.477***\\
 & (0.146) & (1.314) & (1.334)\\
Fixed probabilities framing &  &  & -0.570***\\
 &  &  & (0.195)\\
Female/other & -0.369 &  & -0.373\\
 & (0.270) &  & (0.290)\\
Years since undergraduate degree & -0.089 &  & -0.070\\
 & (0.077) &  & (0.087)\\
Undergraduate region: Asia & -0.472* &  & -0.495*\\
 & (0.229) &  & (0.230)\\
Undergraduate region: Europe & -0.207 &  & -0.160\\
 & (0.273) &  & (0.264)\\
Undergraduate region: missing & 0.013 &  & -0.136\\
 & (0.413) &  & (0.378)\\
Undergraduate region: other & 0.201 &  & 0.178\\
 & (0.658) &  & (0.651)\\
Years since undergraduate degree missing & -0.440 &  & -0.304\\
 & (0.339) &  & (0.293)\\
Log citation count &  & 0.046 & 0.084\\
 &  & (0.106) & (0.113)\\
Workplace type: academic &  & 0.087 & 0.176\\
 &  & (0.174) & (0.177)\\
Accepted at NeurIPS 2018 &  & 0.833 & 0.792\\
 &  & (1.239) & (1.234)\\
Accepted at ICML 2018 &  & 0.563 & 0.447\\
 &  & (1.240) & (1.221)\\
Familiarity with AI safety &  & 0.140 & 0.145\\
 &  & (0.107) & (0.107)\\
Expected value of perceived effect of human-level machine intelligence &  &  & -0.028\\
 &  &  & (0.123)\\
Log citation count missing &  & 0.023 & 0.104\\
 &  & (0.211) & (0.209)\\
Familiarity with AI safety missing &  & -0.532* & -0.497\\
 &  & (0.250) & (0.258)\\
\midrule
Number of respondents & 289 & 289 & 289\\
\bottomrule
\multicolumn{4}{l}{\textsuperscript{} * p < 0.05, ** p < 0.01, *** p < 0.01}\\
\end{tabular}
\end{table}

Model 1 regressed the outcome on a set of demographic variables: gender, a proxy for age --- years since undergraduate degree --- and a proxy for nationality --- region where undergraduate degree was completed.\footnote{As noted in Footnote 3 of \citep{Zwetsloot2019}, students often complete their undergraduate degree in the country that they were born in. However, the misclassification rate using this methodology has been found to be up to 20\% \citep{Zwetsloot2019} and so care should be taken when interpreting such results as a proxy for nationality or region of origin.} Individuals who completed their undergraduate studies in Asia predicted significantly shorter human-level machine intelligence timelines in comparison to researchers that completed their undergraduate degrees in North America, a finding that mirrors that of \cite{GraceEtAl}.

Model 2 regressed the outcome on a set of variables that describe the work experience and background of the AI/ML researchers: citation count, whether they work in academia or industry, which conference they were accepted at, their familiarity with AI safety. No predictors except the dummy variable for missing AI safety familiarity data were found to significantly predict the outcome.

Model 3 included the complete set of our predictors comprising those in Models 1 and 2 in addition to question framing (fixed years vs. fixed probabilities) and the expected value of the perceived impact human-level machine intelligence will have on the long-run future of humanity. Here we found that only question framing significantly predicts human-level machine intelligence timelines. Like \cite{GraceEtAl}, we found that the fixed probabilities framing results in shorter timelines than the fixed year framing. 

\subsection{Panel Sample Differences in Aggregate Forecast CDFs between 2016 and 2019}

The results for the 49 respondents surveyed in both 2016 and 2019 initially appear to show a lengthening of the predicted HLMI date from 2062 to 2076. However, randomization inference tests using both a Wasserstein metric and Kolmogorov-Smirnov statistic showed no significant differences in the aggregate median parameter CDFs for the panel sample between 2016 and 2019. No significant mean within-subject differences were found using the Wasserstein metric randomization inference. When both framings are analyzed separately using a clustered Wilcoxon signed rank test\footnote{We performed a clustered Wilcoxon signed rank test using the Rosner-Glynn-Lee method because each respondent provided three data points.}, the fixed years framing results in a \textit{p}-value of $0.03$ when comparing the panel sample results while the fixed probabilities framing results in a \textit{p}-value of $0.86$. 

\subsection{Impact of Human/High-Level Machine Intelligence}

% Here too, the panel sample was asked about \textit{high}-level machine intelligence while the cross-sectional sample was asked about \textit{human}-level machine intelligence. Since a robust linear model regression analysis found no significant difference in the expected value\footnote{The expected value for each respondent is calculated by multiplying each of the probability estimates by the scale number assigned to the outcome and summing together these values. In mathematical notation this can be represented as \begin{math}\textup{E[}\mathit{X}]= \sum_{i=1}^{k}x_{i}p_{i}\end{math}, such that the expected value is the weighted sum of the scale values,\begin{math}x_{i}\end{math}, where the probabilities,\begin{math}p_{i}\end{math}, are the assigned weights.} of the impact of human/high-level machine intelligence across these two definitions, we collapsed across the 2019 responses of the cross-sectional and panel samples in subsequent analyses for a total sample of $373$ respondents. Below, we will refer to human/high-level machine intelligence as HLMI. 

% Figure \ref{fig:hlmiimpact-expectedvalue} shows the positive beliefs of AI/ML researchers. It plots the density distribution of the expected value of HLMI impact across all respondents. 

\begin{figure}[H]
    \centering
    \includegraphics[width=0.85\textwidth]{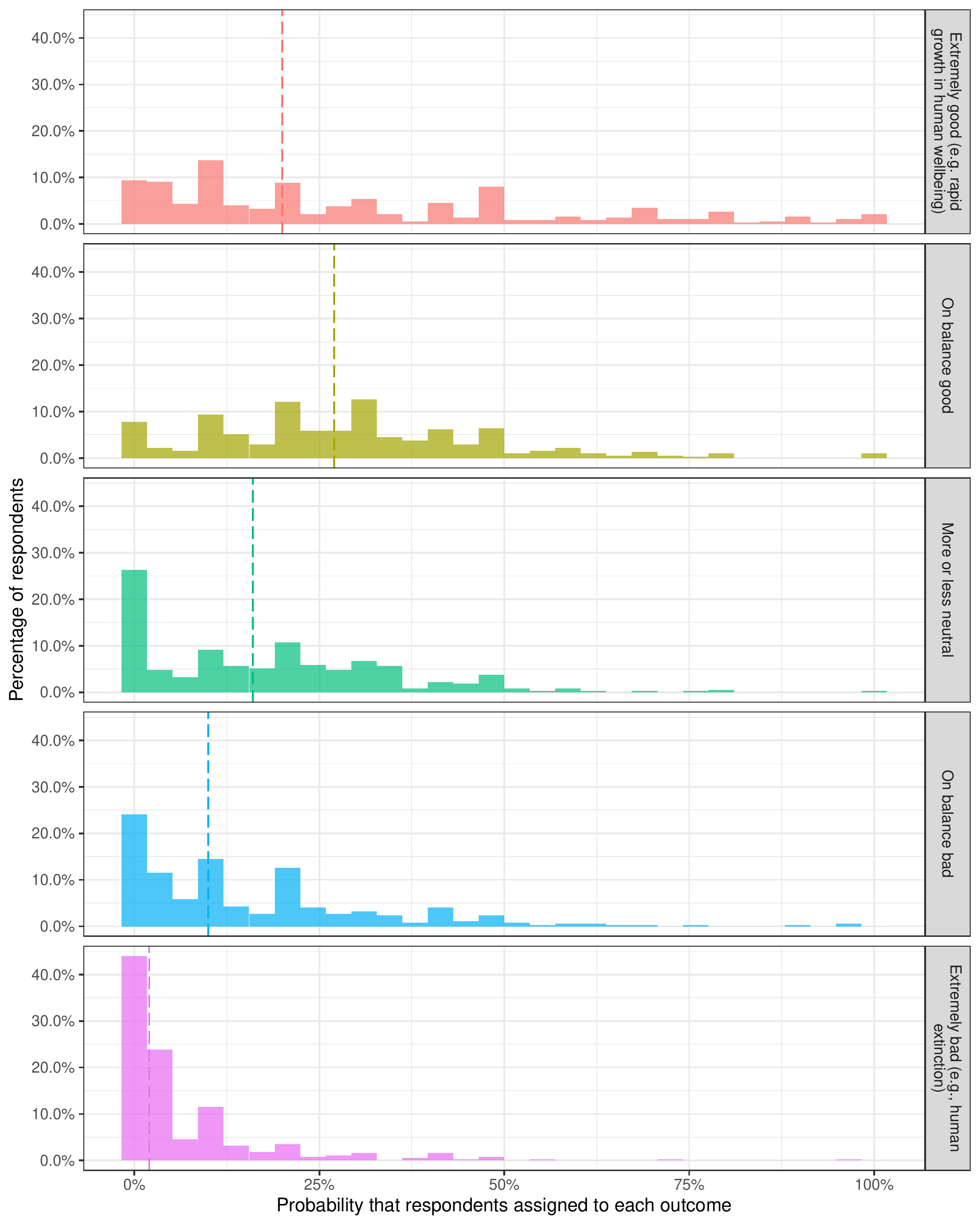}
    \caption{Distribution of respondents' assigned probabilities to each outcome when asked about the societal impact of human/high-level machine intelligence. Respondents were given the five outcomes and asked to assign a probability to each outcome. The survey tool compelled the respondents such that the probabilities summed up to 100\%. 373 respondents completed this question. The dashed lines represent the median probability assigned to each outcome.}
    \label{fig:hlmi_histogram}
\end{figure}

Combining the probabilities of both bad and both good outcomes revealed that AI/ML researchers' probabilities of the effect of AI on humanity in the long run skews optimistic, as seen in Fig. \ref{fig:hlmi_histogram}. Interestingly, with 11\% of respondents ascribing no probability to the possibility that HLMI could have a ``more or less neutral'', ``on balance bad'', or ``extremely bad'' outcome on humanity, and a 100\% chance that the impact of HLMI will be ``on balance good'' or ``extremely good.'' This could illustrate a degree of overconfidence in the positive impact of HLMI.

\begin{figure}[H]
    \centering
    \includegraphics[width=\textwidth]{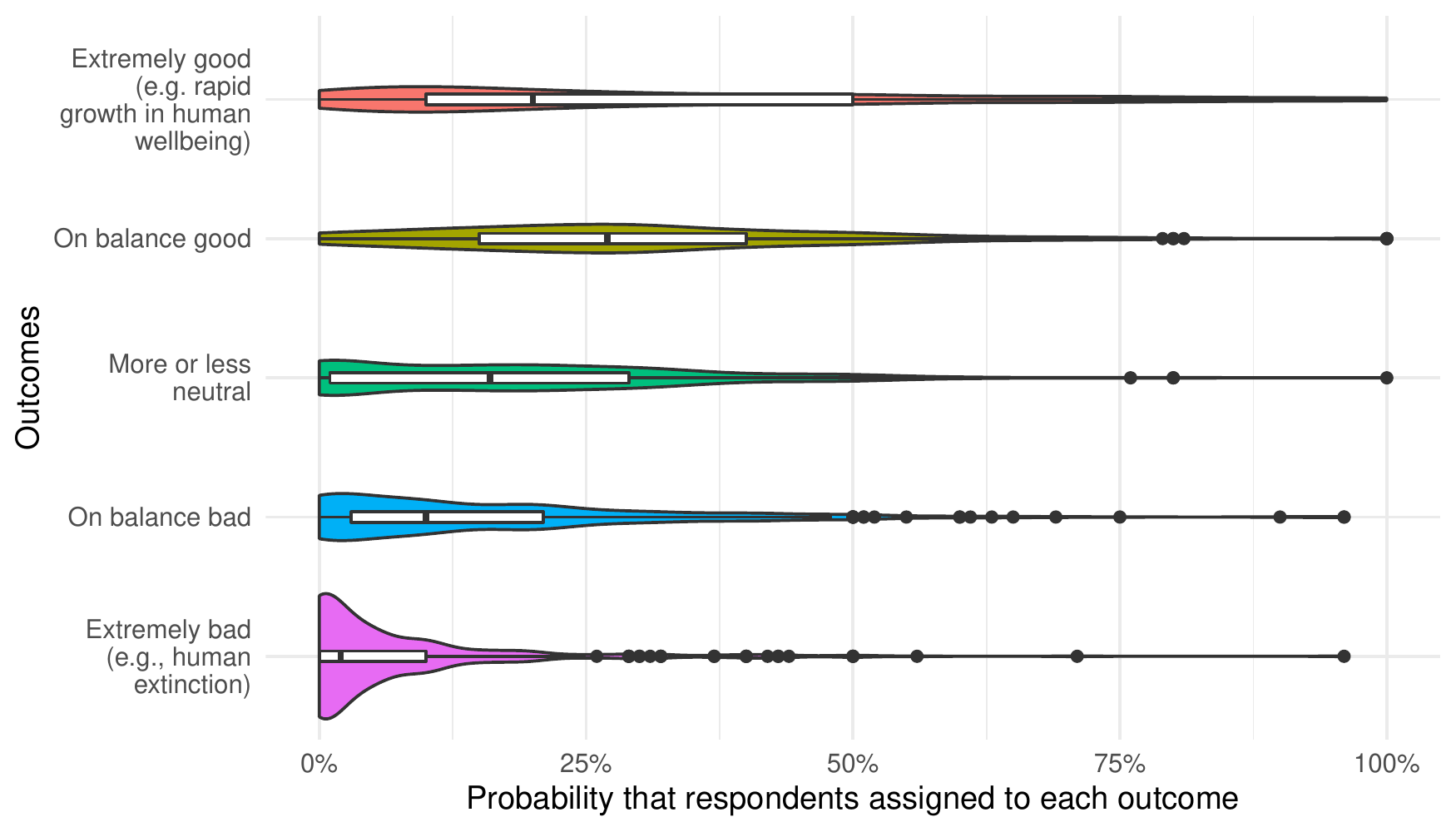}
    \caption{Violin plot --- a mirrored density distribution ---  of the cross-sectional and panel sample probability responses in 2019 for each answer option for the potential impact HLMI could have on humanity in the long run. The first, second (median), and third quartiles of the probability responses are indicated.}
    \label{fig:hlmiimpactviolinplot}
\end{figure}

Mirrored density plots of each response outcome in Fig. \ref{fig:hlmiimpactviolinplot} show that there still exists a certain degree of fear, though, even amongst those working on developing these technologies, that HLMI could have an overall negative impact on humanity. In the 2019 survey, the median respondent ascribed a probability of 10\% that HLMI would be ``on balance bad'' for humanity in the long run, with 2\% noting it could lead to human extinction, and 16\% noted that it would be ``more or less neutral.''\footnote{The corresponding mean results were 16\% ``on balance bad'', 7\% ``extremely bad'' and 18\% neutral. \cite{GraceEtAl} found that the median respondent placed similar probabilities of 20\%, 10\%, and 5\% on the same outcomes, from ``neutral'' to ``extremely bad.''}

\subsection{Differences in Milestone Predictions in 2016 and 2019}

Randomization inference tests did not reveal statistically significant differences (at the 5\%-level) between the survey samples in terms of individual milestones except for three milestones (group unseen objects, perform well in the Putnam Competition, and win at Atari). Nevertheless, this result only held for tests using the Wasserstein metric and not the Kolmogorov-Smirnov statistic (see Tables \ref{tab:cross-ri-ws}-\ref{tab:cross-ri-ks-1}). Furthermore, we tested for differences between the two samples across the milestones using nonparametric combination (NPC) tests \citep{caughey2017nonparametric}, which is used to jointly test hypotheses individually tested using randomization inference. Again, the NPC tests were inconclusive (Table \ref{tab:npc-results}): while the test using the Wasserstein metric suggested there was a statistically significant difference (at the 1\%-level) between the two samples, the tests using the Kolmogorov-Smirnov statistic did not.

For the panel sample, across the milestones, the 2016 and 2019 forecasts were not statistically different (Table \ref{tab:panel-ri}). We tested the forecasts as a whole rather by milestone because all of the milestones had fewer than 10 pairs of forecasts. Analyzing the raw data using the clustered Wilcoxon signed rank test, shown in Table \ref{tab:wilcoxon-results} we found that for the fixed probabilities framing, the forecasted milestone timelines in 2019 were longer than the forecasted milestone timelines in 2016 ($p$-value $= 0.041$).\footnote{We were unable to test the raw data for the fixed years group because we changed the year anchors from 10, 20, and 50 years in 2016 to 1, 5, and 10 years in 2019.}

\subsection{Correlation Between Human-level Machine Intelligence and Milestone Forecasts}

In the cross-sectional sample, for the milestone forecasts, there was no statistically significant correlation with the human-level machine intelligence forecasts. We tested this by using the log year of achieving human-level machine intelligence with 50\% probability to predict the log year of achieving the milestone with 50\% probability. In one model, we used fixed effects to account for variations in milestones; in another model, we used random effects instead. In both models, we clustered the standard errors by respondent. In both models (Table \ref{tab:hlmi-milestone-regression}), we find no correlation between the log year of achieving human-level machine intelligence with 50\% probability and the log year of achieving the milestones with 50\% probability ($p$-value $= 0.695$ in the fixed effects model; $p$-value $= 0.785$ in the random effects model). 

\subsection{Potential Milestone Feasibility and Resolutions}

A number of the milestones have been close to resolution or might be already be feasible. For example, DeepMind's AlphaStar did indeed win against top human players in StarCraft, placing it in the top 0.15\% (or roughly 1,000 players) on the European StarCraft server, while using video input, it did not beat the \textit{very best} human players \citep{vinyals_grandmaster_2019}. In addition, many of the milestones predicted in 2016 and 2019 referred to natural language processing (NLP), a field which appears to have seen significant progress since 2016 with the publication of for example BERT in 2018 \citep{devlin-etal-2019-bert}, GPT-3 in 2020 \citep{GPT-3}, and Gopher in 2021 \citep{Gopher}. For example, respondents produced forecasts on when AI systems would be able to write a high-school history essay that would receive high grades, answer any ``easily Googleable'' factoid better than a human, or read a passage aloud in a manner indistinguishable from a voice actor. Given recent progress in the field, these milestones may be feasible to meet today, but to our knowledge, no one has tested recent NLP models against these specific benchmarks. 

Our AngryBirds milestone\footnote{Play new levels of Angry Birds better than the best human players. Angry Birds is a game where players try to efficiently destroy 2D block towers with a catapult. For context, this is the goal of the International Joint Conference on Artificial Intelligence Angry Birds AI Competition \citep{angryBirdsCompetition}.} presents a similar case. In 2016, respondents provided an aggregate prediction of 50\% chance of the AngryBirds milestone being feasible by 2010.6 and a 75\% chance it would be feasible by 2024.1. In 2019 our respondents forecast an aggregate of 50\% probability by 2022.8 and 75\% probability by 2024.1. Nonetheless, the milestone is yet to be achieved. It may be that the milestone is feasible, but no large actor has yet put effort into solving it. To date, academic labs --- which typically have less computational resources --- have won the competition most years. The organizers of the conference speculate that solving the challenge is not yet feasible --- that it is yet to be solved because the deep learning is not well-suited to the task as the systems only have access to video data \citep{AngryBirds}.

\newpage

\section{Supplementary Text: Survey Text}

Our 2019 survey asked respondents about their views on a variety of topics from the ethics and governance of AI to immigration and AI progress. This paper presents the results for the forecasts AI/ML researchers made in regard to human/high-level machine intelligence, AI milestones, and the impacts of human/high-level machine intelligence on humanity in the long-run. Results for the other survey questions can be found elsewhere \citep{Zhang2021, skilledAndMobile2021}.

\subsection{HLMI Forecasts}

Similar to several previous surveys of AI experts \citep{GraceEtAl,gruetzemacher2019forecasting,MullerBostrom}, we asked our respondents to make forecasts about a form of HLMI. For our cross-sectional sample we defined this as:

\begin{quote}
Human-level machine intelligence (HLMI) is reached when machines are collectively able to perform almost all tasks (>90\% of all tasks) that are economically relevant* better than the median human paid to do that task in 2019. You should ignore tasks that are legally or culturally restricted to humans, such as serving on a jury.
*We define these tasks as all the ones included in the Occupational Information Network (O*NET) dataset. O*NET is a widely used dataset of tasks required for current occupations.
\end{quote}
For our panel sample we retained the definition used by \cite{GraceEtAl} to be able to compare responses made by respondents in 2016 and 2019:
\begin{quote}
Say we have `high-level machine intelligence' when unaided machines can accomplish every task better and more cheaply than human workers. Ignore aspects of tasks for which being a human is intrinsically advantageous, e.g. being accepted as a jury member. Think feasibility, not adoption.
For the purposes of this question, assume that human scientific activity continues without major negative disruption.
\end{quote}

This question was asked with two different framings: In the fixed years framing, cross-sectional sample respondents were asked, ``How likely is it that HLMI exists in 10 years? 20 years? 50 years?'' and responded with a probability value for each (three total). Panel sample respondents were asked, ``How likely is it that HLMI exists in 10 years? 20 years? 40 years?'', to mirror the question asked in the 2016 survey. In the fixed probabilities framing, respondents were asked, ``How many years from now will HLMI exist with 10\% probability? 50\% probability? 90\% probability?'' and responded with a year value for each. These two framings are known to lead to significantly different forecasts \citep{GraceEtAl}.

\subsection{AI Milestones}

We used also used fixed years and fixed probabilities framings to ask researchers for their predictions on when certain AI milestones would be ``achieved and made public'' . In the fixed years framing respondents were asked:

\begin{quote}
How likely is it that the following AI progress milestone is achieved and made public in 1 year? 5 years? 10 years?
\end{quote}

In the fixed probabilities framing respondents were asked:

\begin{quote}
How many years from now do you think that the following AI progress milestones will be achieved and made public with 10\% probability? 50\% probability? 90\% probability? 
\end{quote}

Some key differences exist in comparison to the 2016 survey in the way milestone predictions were elicited \citep{GraceEtAl}. Firstly, the \cite{GraceEtAl} survey asked when a task was ``feasible''. Feasible was defined as ``if one of the best resourced labs could implement it in less than a year if they chose to. Ignore the question of whether they would choose to.'' Secondly, while the fixed probabilities framing probabilities used were the same, the fixed years framing were different: 10, 20, or 50 years. Finally, while there is overlap between the milestones used, some use slightly different descriptions.

The milestone descriptions used in the current survey are shown below, while the 2016 ones can be found in the appendix of \cite{GraceEtAl}:

\textbf{Phone banking services}

Provide phone banking services as well as human operators can, without annoying customers more than humans. This includes many one-off tasks, such as helping to order a replacement bank card or clarifying how to use part of the bank website to a customer. Provide phone banking services as well as human operators can, without annoying customers more than humans. This includes many one-off tasks, such as helping to order a replacement bank card or clarifying how to use part of the bank website to a customer.

\textbf{One-shot learning}

One-shot learning: given only one labeled image of a new object, recognize the object in real-world scenes as well as a typical human can (i.e., including in a wide variety of scenes). For example, given only one image of a platypus, recognizes platypuses in nature photos. The system may train on labeled images of other objects.

\textbf{Transcribe human speech}

Transcribe human speech with a variety of accents in a noisy environment as well as a typical human can.

\textbf{Text to speech voice actor
}

Take a written passage and output a recording that can’t be distinguished from a voice actor, by an expert listener.

\textbf{Prove math theorems}

Routinely and autonomously prove mathematical theorems that are publishable in top mathematics journals today, including generating the theorems to prove.

\textbf{Perform well in Putnam}

Perform as well as the best human entrants in the Putnam competition—a math contest whose questions have known solutions, but which are difficult for the best young mathematicians.

\textbf{Win at Angry Birds}

Outperform the best human players at the Angry Birds AI Competition’s Man vs Machine Challenge. Angry Birds is a video game where players try to destroy 2D block towers with a catapult efficiently. For context, this is the goal of the IJCAI Angry Birds AI competition.  Reference: aibirds.org

\textbf{Win at Atari}

Outperform professional game testers on all Atari games, with the AI system having learned all games from scratch (i.e., without human demonstrations). This includes games like Frostbite, which require planning to achieve sub-goals and have posed problems for deep Q-networks.

\textbf{Fold laundry}

Fold laundry as well and as fast as the median human clothing store employee.

\textbf{Write python code}

Write concise, efficient, human-readable Python code to implement simple algorithms like Quicksort. That is, the system should write code that sorts a list, rather than just being able to sort lists. Suppose the system is given only a specification of what counts as a sorted list and several examples of lists undergoing sorting by Quicksort.

\textbf{Write history essay}

Write a 10-page essay for a college history class that would receive high grades and pass plagiarism detectors. For example, answer a question like ``How did the industrial revolution impact farming in the United Kingdom?''.

\textbf{Compose top-40 song}

Compose a song that reaches the US Top 40. The system should output the complete song as an audio file.

\textbf{Write NYT bestseller}

Write a novel that reaches the New York Times best-seller list.

\textbf{Explain moves in computer game}

For any computer game that can be played well by a machine, explain the choice of moves in a way that helps a novice human to improve at the game.

\textbf{Output virtual world equations}

After spending time in a virtual world, output the differential equations governing that world in symbolic form. For example, the agent is placed in a game engine where Newtonian mechanics holds and the agent is then able to conduct experiments with a ball and output Newton’s laws of motion.

\textbf{Win hacking competition}

Fully autonomous system can win against the best human team in a DEFCON Capture the Flag finals – ``widely considered the world cup of hacking''  – where the human team is not making substantial use of AI tools themselves.

\textbf{Find software vulnerability}

A serious software vulnerability is found primarily by a machine learning system reading the software's code. The software must be widely used.

\textbf{Human-level translation}

Perform translation as well as a human who is fluent in both languages but unskilled at translation, for most types of text, and for most popular languages (including languages that are known to be difficult, like Czech, Chinese and Arabic).

\textbf{Group unseen objects}

Correctly group images of previously unseen objects into classes, after training on a similar labeled dataset containing completely different classes. The classes should be similar to the ImageNet classes.

\textbf{Assemble LEGO}

Physically assemble any LEGO set given the pieces and instructions, using specialized robotics hardware. For context, Fu et al. (2016) successfully joins single large LEGO pieces using model-based reinforcement learning and online adaptation. 

\textbf{Autonomous vehicles drive best}

An autonomous vehicle can drive better than the median human across a wide range of driving conditions. ``Driving better'' means driving more safely and efficiently.

\textbf{Assemble IKEA}

A fully autonomous robot correctly assembles an IKEA chair faster than the median human, given the standard instructions manual.

\textbf{Beat human game speedruns}

Surpassing human performance at speed runs – i.e. completing the game as quickly as possible – in open-world video games (such as games in the Grand Theft Auto or Minecraft).

\subsection{HLMI Impact}

In both 2016 survey and in the current survey from 2019, respondents were asked:

\begin{quote}
Assume for the purpose of this question that HLMI will at some point exist. How positive or negative do you expect the overall impact of this to be for humanity, in the long run?  

Please answer by saying how probable you find the following kinds of impact, with probabilities adding to 100\%: 

\begin{itemize}
    \item Extremely good (e.g., rapid growth in human flourishing) (2)
    \item On balance good (1)
    \item More or less neutral (0)
    \item On balance bad (-1)
    \item Extremely bad (e.g., human extinction) (-2)
\end{itemize}
\end{quote}

In addition, in 2019, the definition for human-level machine intelligence was shown. In the 2016 survey respondents were reminded before the question:

\begin{quote}
The following questions ask about `high–level machine intelligence' (HLMI). Say we have `high-level machine intelligence' when unaided machines can accomplish every task better and more cheaply than human workers. Ignore aspects of tasks for which being a human is intrinsically advantageous, e.g. being accepted as a jury member. Think feasibility, not adoption. 
\end{quote}

\newpage

\section{Additional Figures}

\subsubsection{High-level machine intelligence forecasts by demographic subgroups}

\begin{figure}[ht]
\caption{Forecasts of human-level machine intelligence by demographic subgroups (Part 1). The CDFs were produced using the median parameters aggregation method.}
\label{fig:forecasts-hlmi-demo}
\begin{subfigure}{.5\textwidth}
  \centering
  % include second image
  \includegraphics[width=\linewidth]{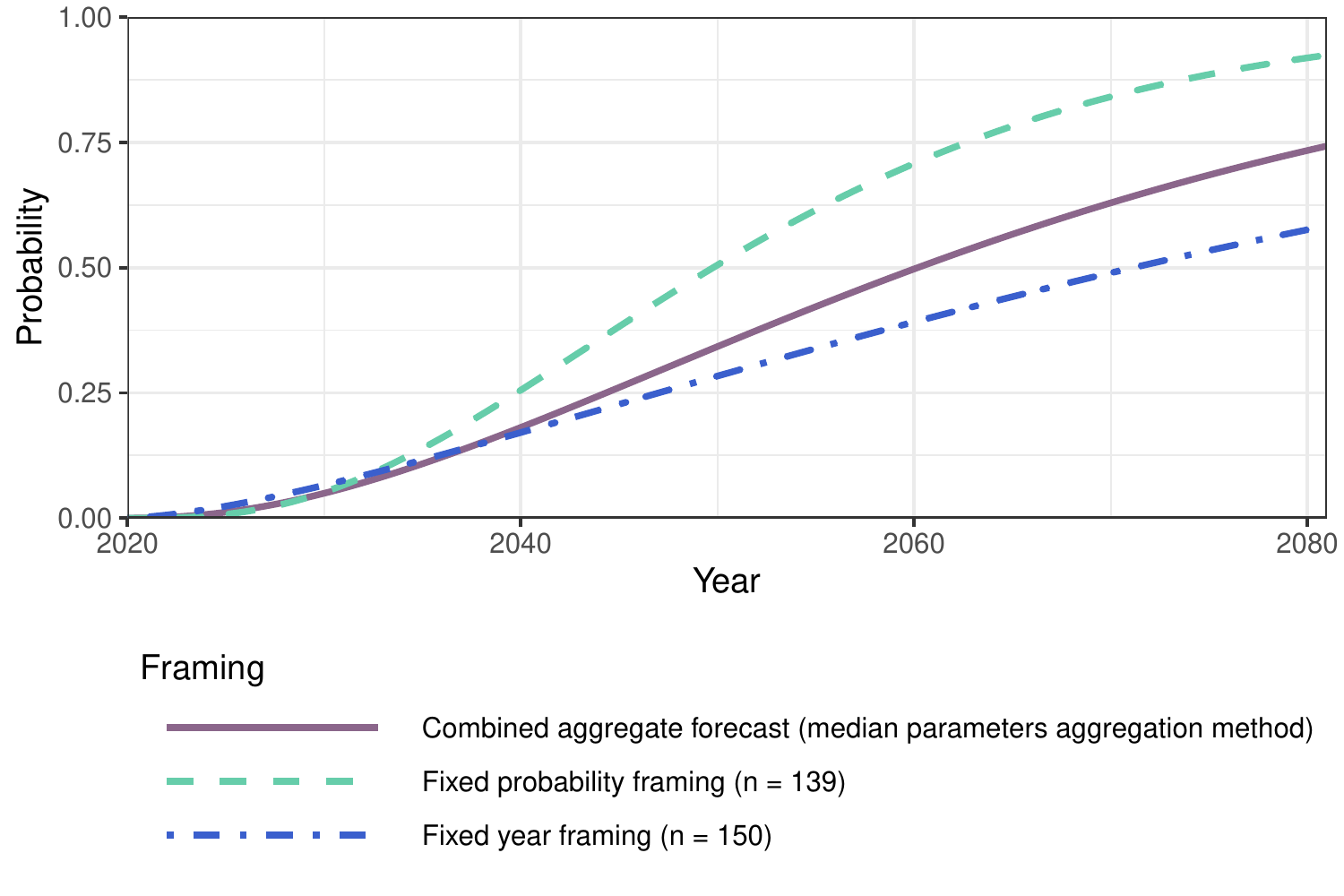}  
  \caption{Framing}
  \label{fig:hlmi-framing}
\end{subfigure}
\begin{subfigure}{.5\textwidth}
  \centering
  % include first image
  \includegraphics[width=\linewidth]{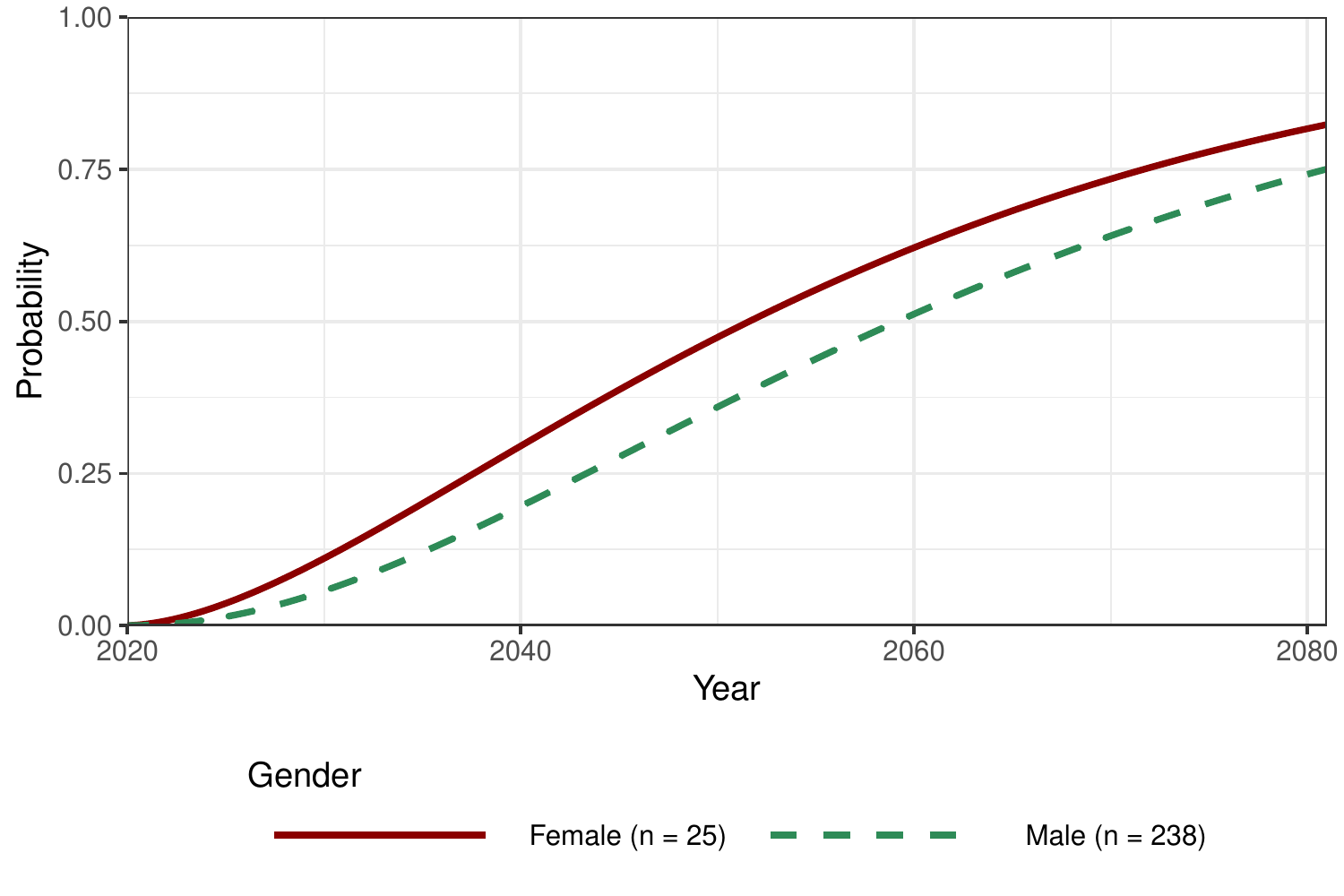}  
  \caption{Gender}
  \label{fig:hlmi-gender}
\end{subfigure}

%\newline

\begin{subfigure}{.5\textwidth}
  \centering
  % include third image
  \includegraphics[width=\linewidth]{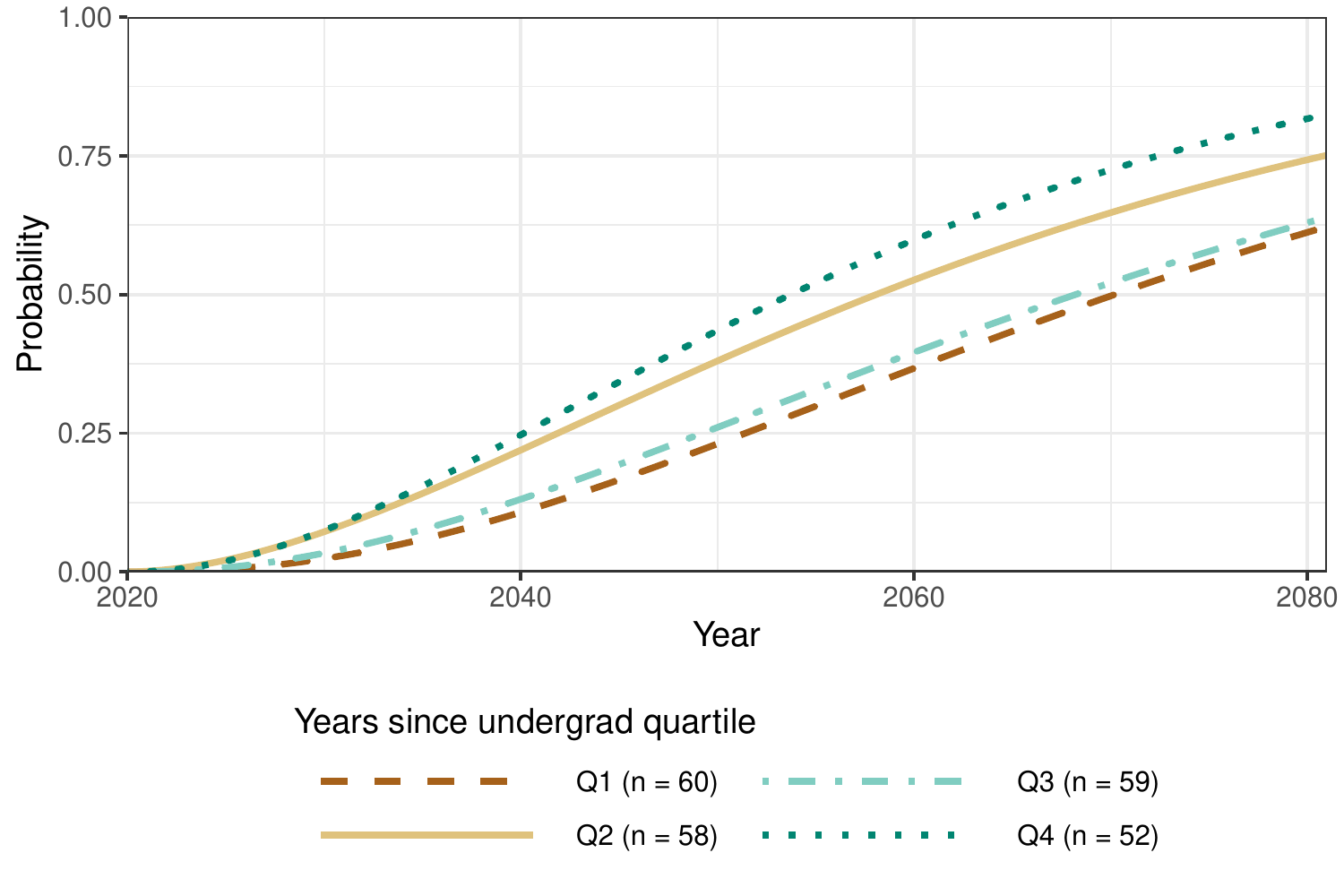}  
  \caption{Years since undergraduate degree}
  \label{fig:hlmi-years-since-undergrad}
\end{subfigure}
\begin{subfigure}{.5\textwidth}
  \centering
  % include third image
  \includegraphics[width=\linewidth]{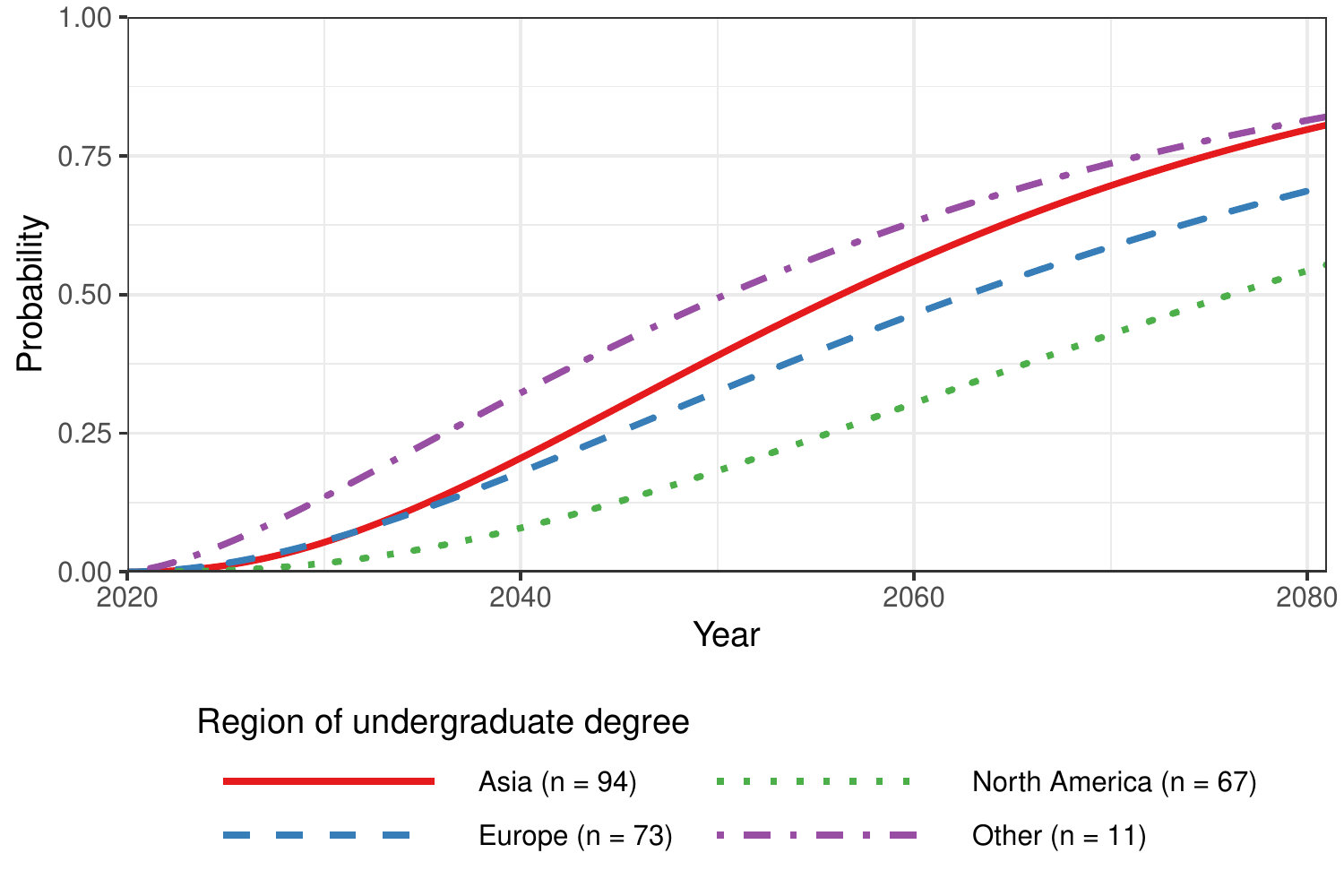}  
  \caption{Undergraduate region}
  \label{fig:hlmi-region}
\end{subfigure}

%\newline

\begin{subfigure}{.5\textwidth}
  \centering
  % include fourth image
  \includegraphics[width=\linewidth]{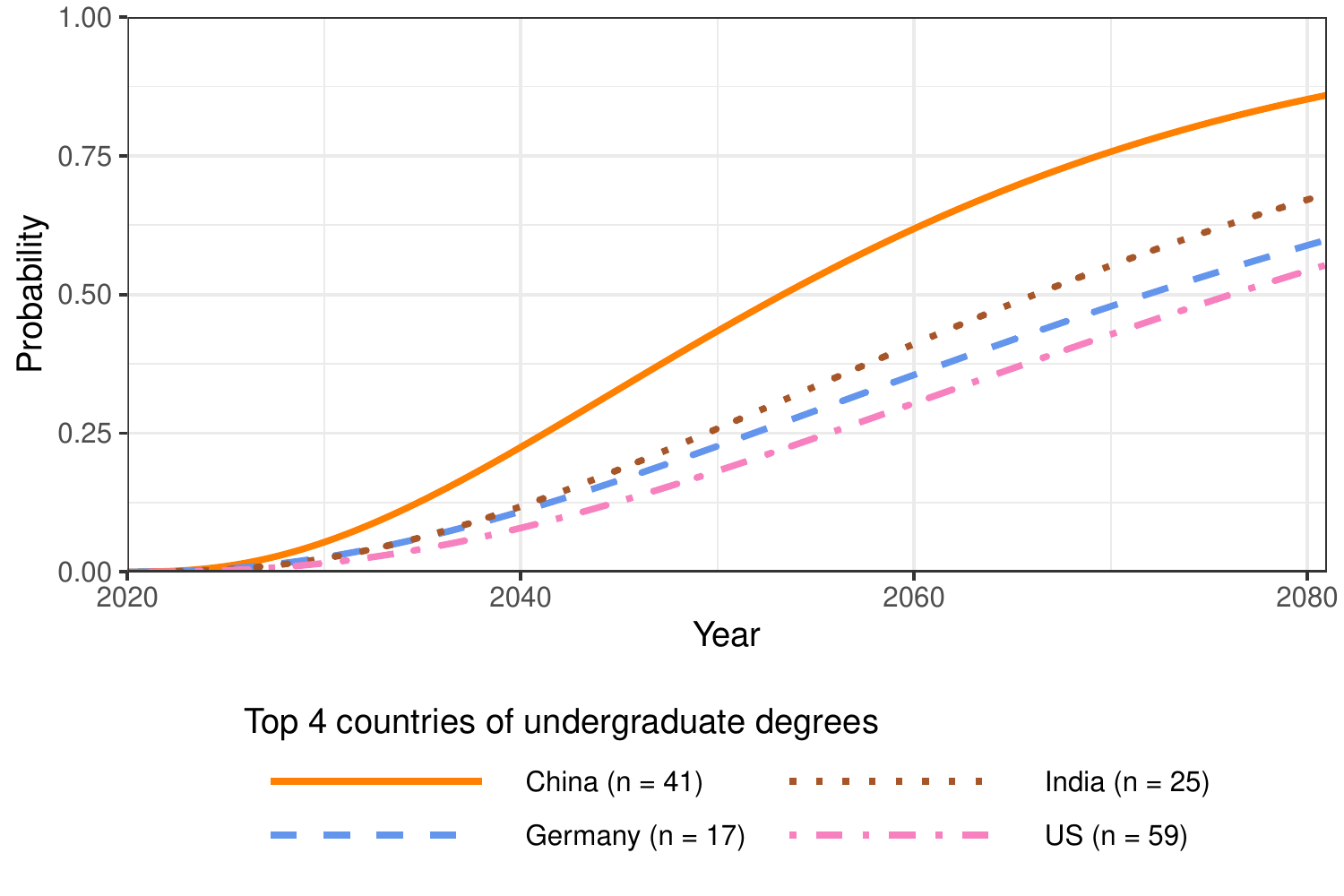}  
  \caption{Undergraduate country (top 4)}
  \label{fig:hlmi-countries}
\end{subfigure}
\begin{subfigure}{.5\textwidth}
  \centering
  % include fourth image
  \includegraphics[width=\linewidth]{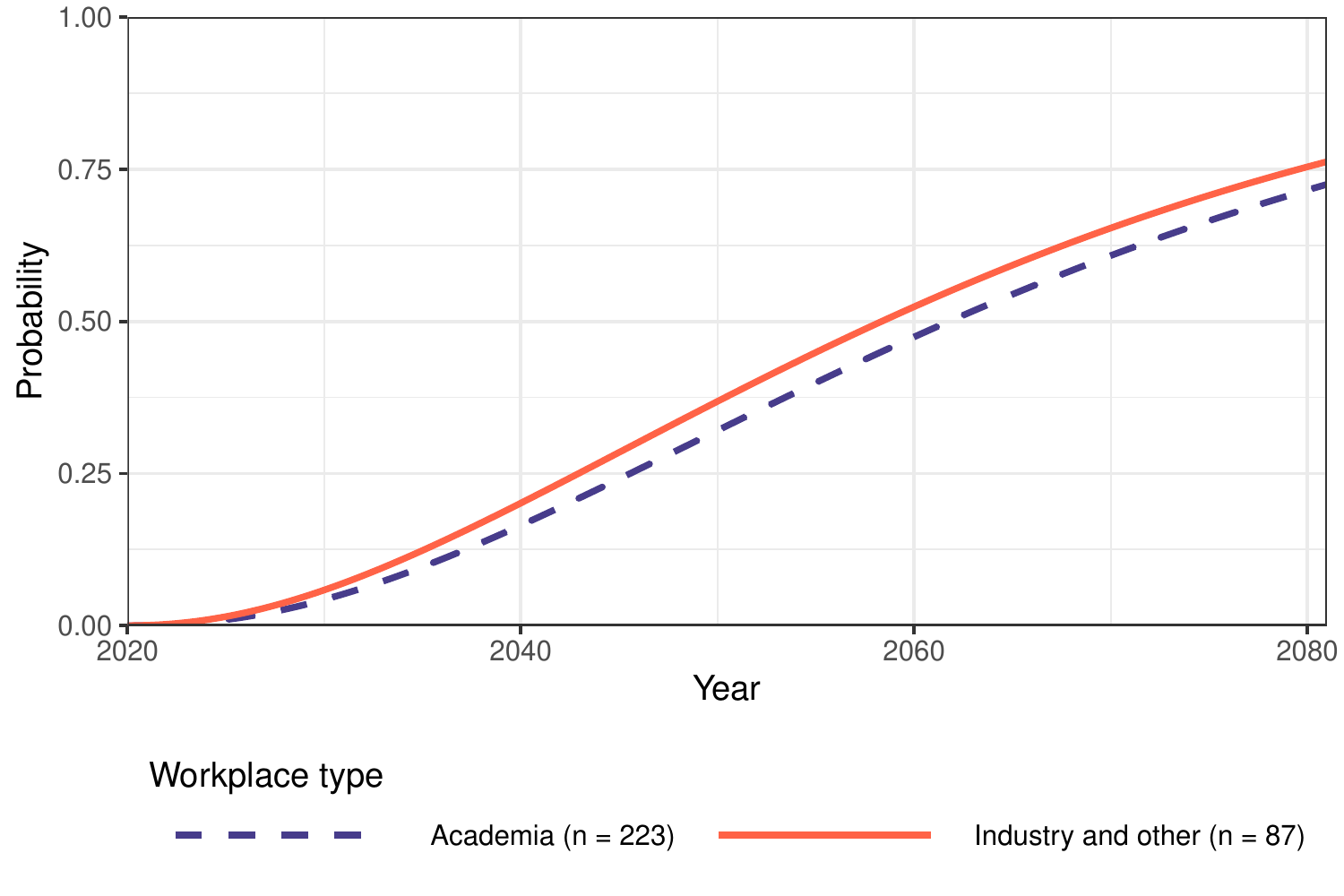}  
  \caption{Workplace type}
  \label{fig:hlmi-workplace}
\end{subfigure}
\end{figure}

\clearpage

\begin{figure}[H]
    \ContinuedFloat % continue from previous page
\begin{subfigure}{.5\textwidth}
  \centering
  % include third image
  \includegraphics[width=\linewidth]{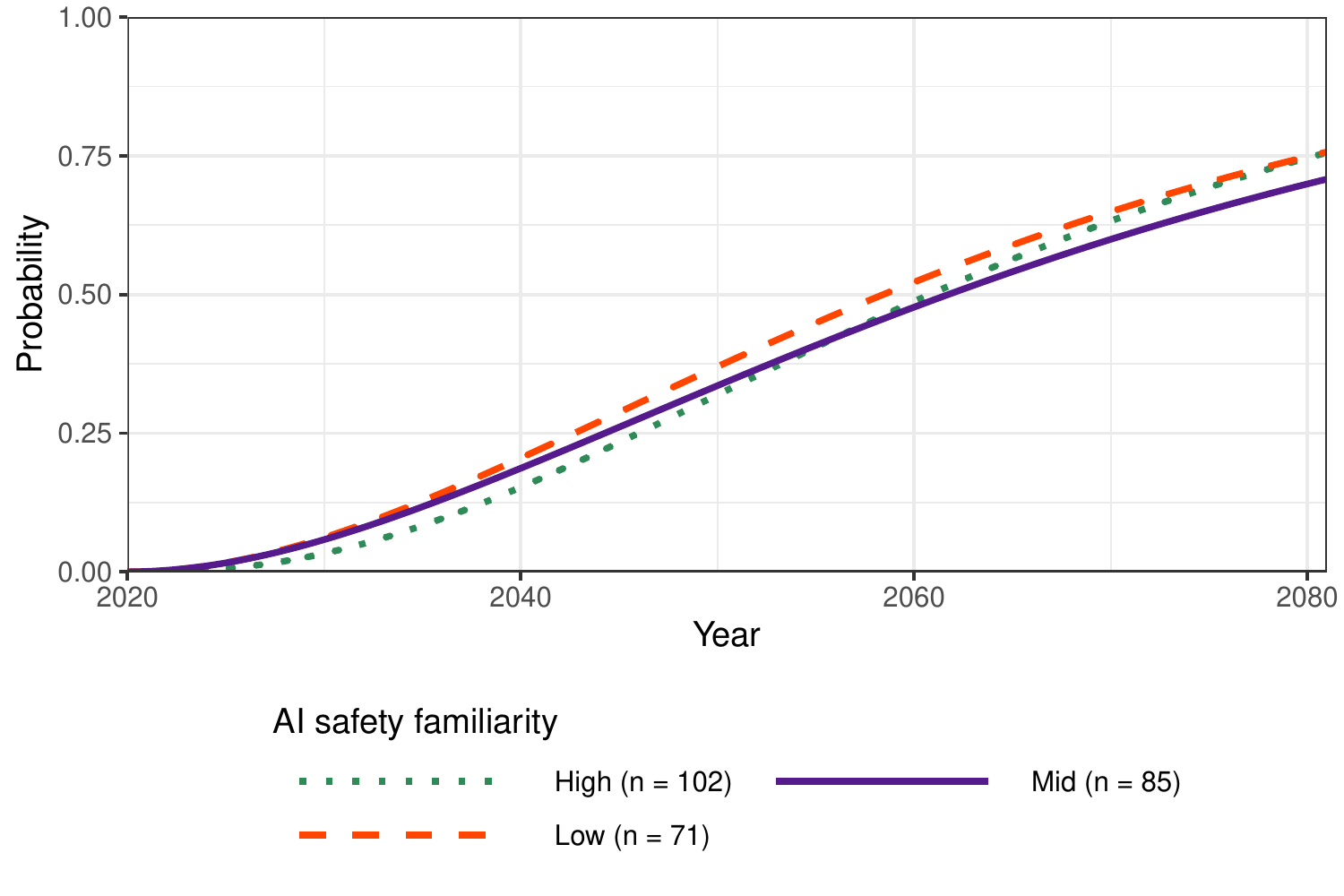}  
  \caption{Familiarity with AI safety}
  \label{fig:hlmi-familiarity-ai-safety}
\end{subfigure}
\begin{subfigure}{.5\textwidth}
  \centering
  % include fourth image
  \includegraphics[width=\linewidth]{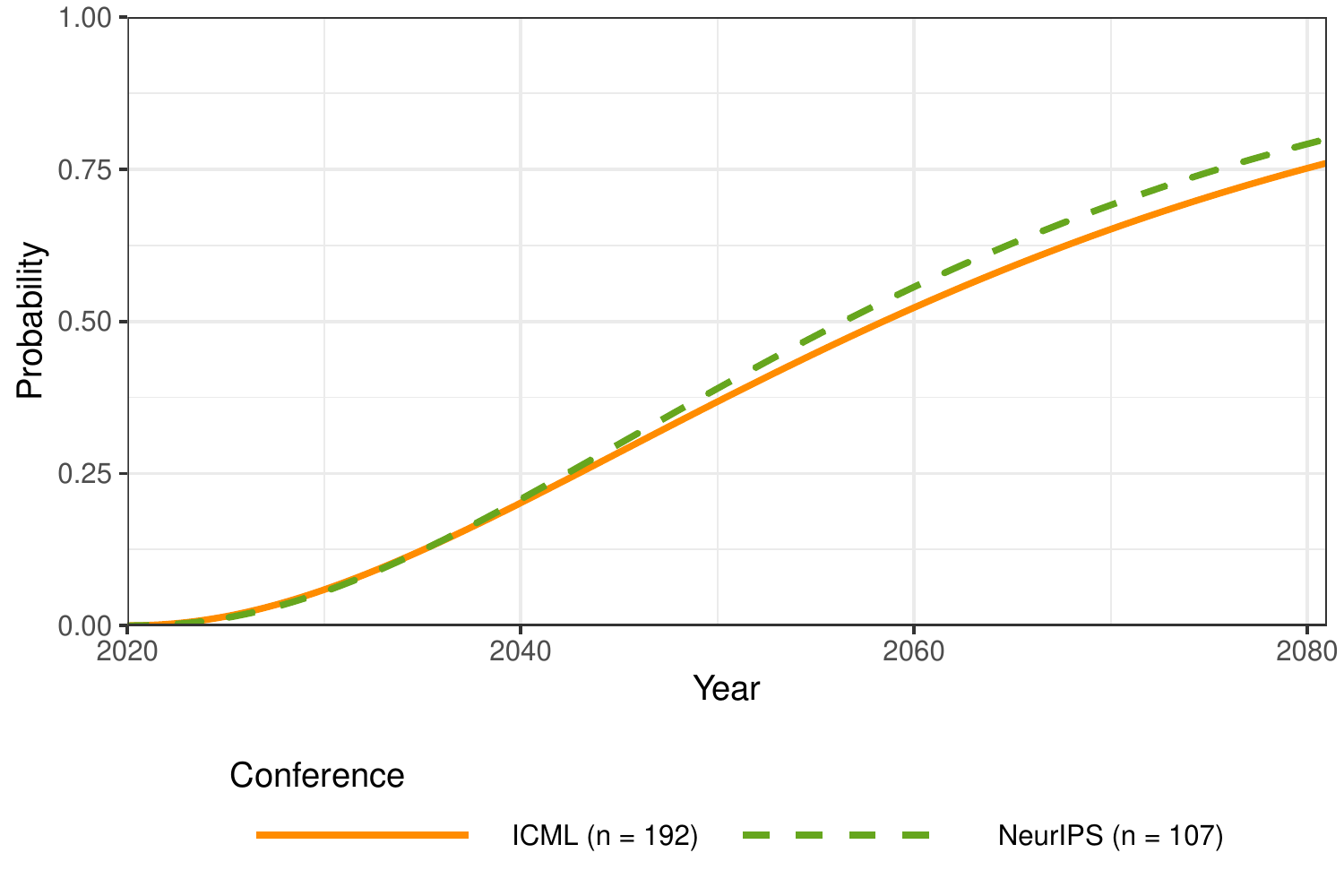}  
  \caption{Which conference the researcher published in}
  \label{fig:hlmi-conference}
\end{subfigure}

%\newline

\begin{subfigure}{.5\textwidth}
  \centering
  % include fourth image
  \includegraphics[width=\linewidth]{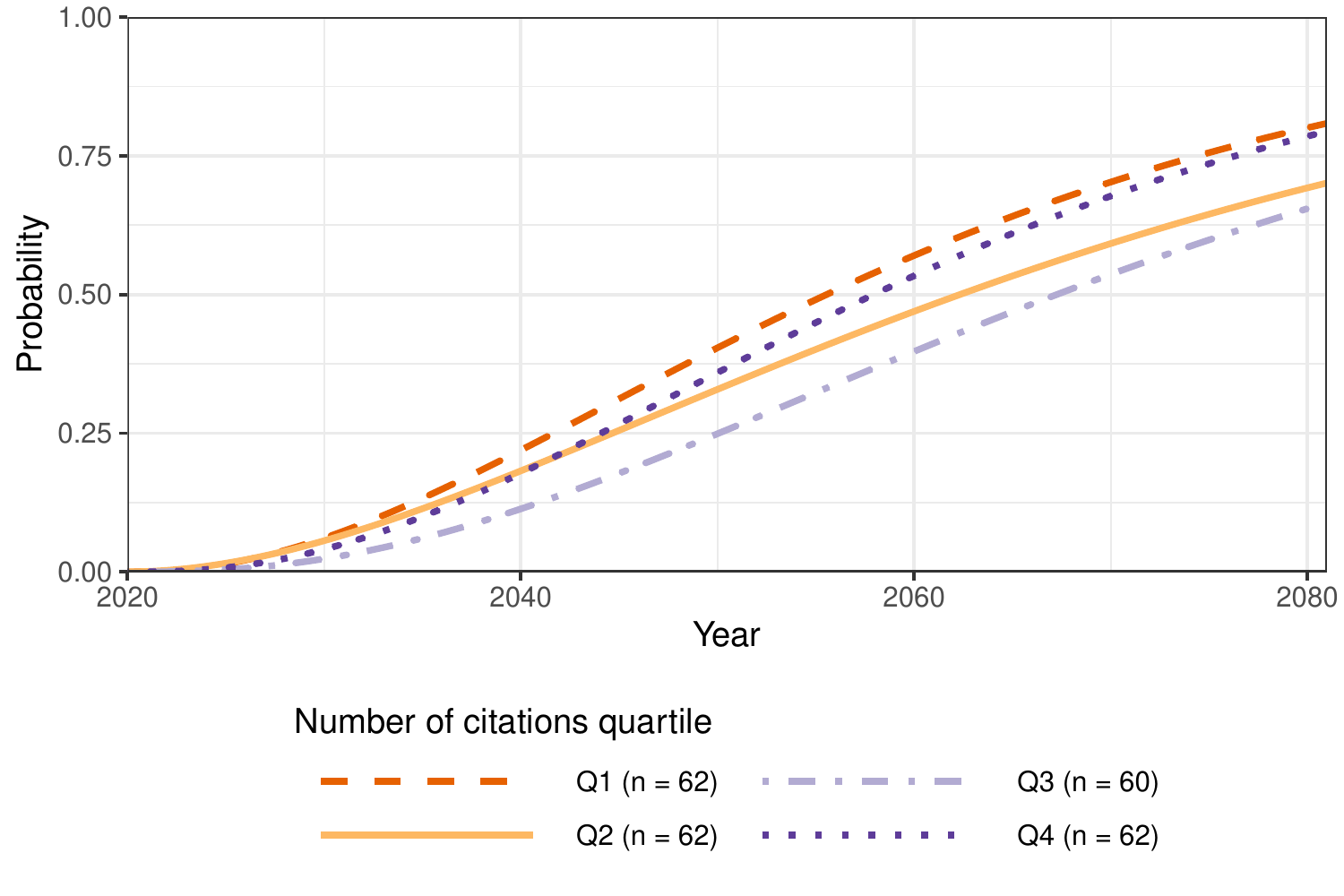}  
  \caption{Number of citations}
  \label{fig:hlmi-citation}
\end{subfigure}
\end{figure}

\FloatBarrier

\subsection{HLMI Forecasts}

\begin{figure}[H]
    \centering
    \includegraphics{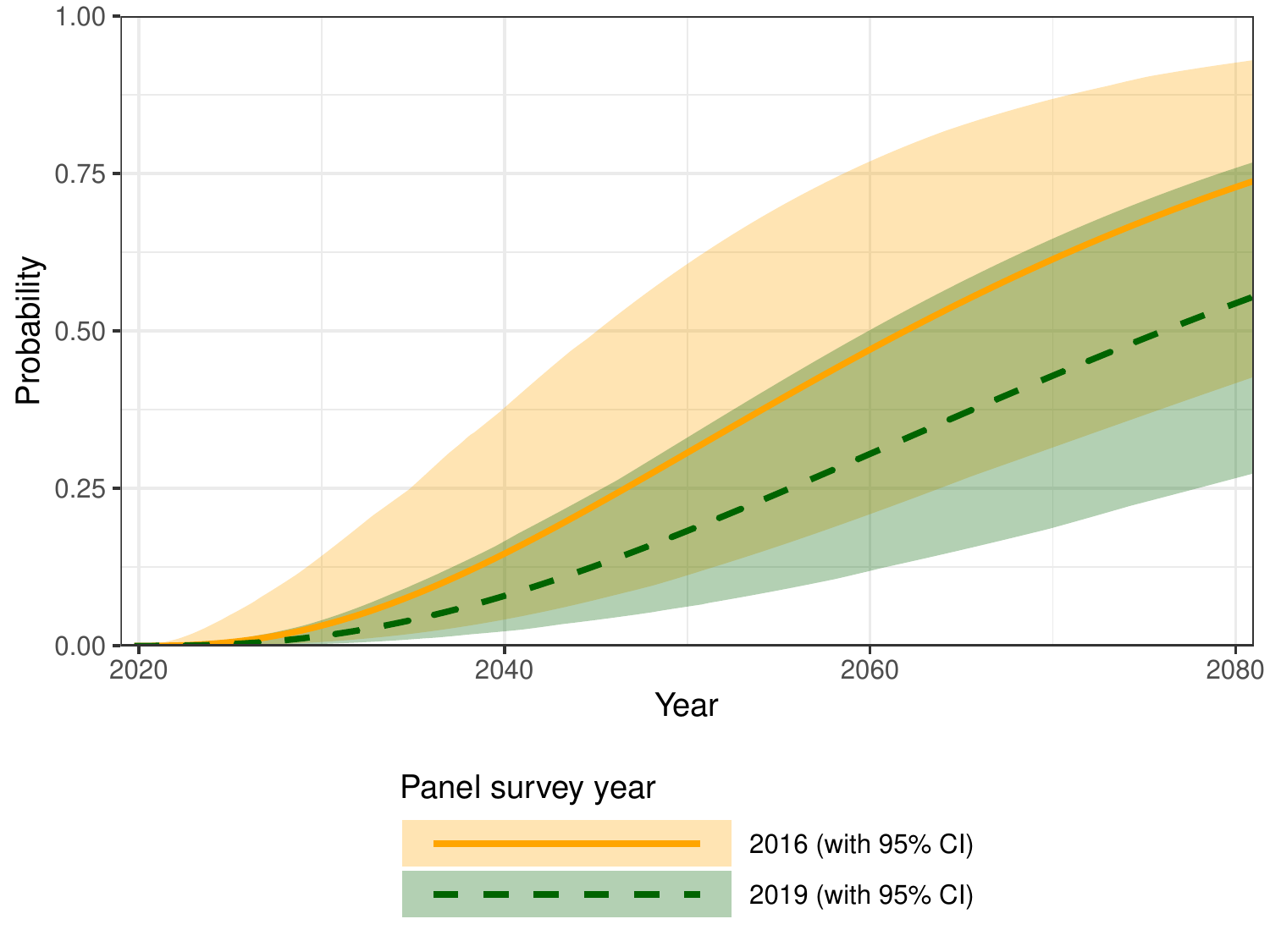}
    \caption{Comparing aggregate forecast CDFs in 2016 and 2019 for the panel sample that were presented with the 2016 definition of high-level machine intelligence.}
    \label{fig:hlmipanel2016v2019}
\end{figure}

\subsection{Impact of HLMI}

\FloatBarrier

\begin{figure}[H]
    \centering
    \includegraphics[width=\textwidth]{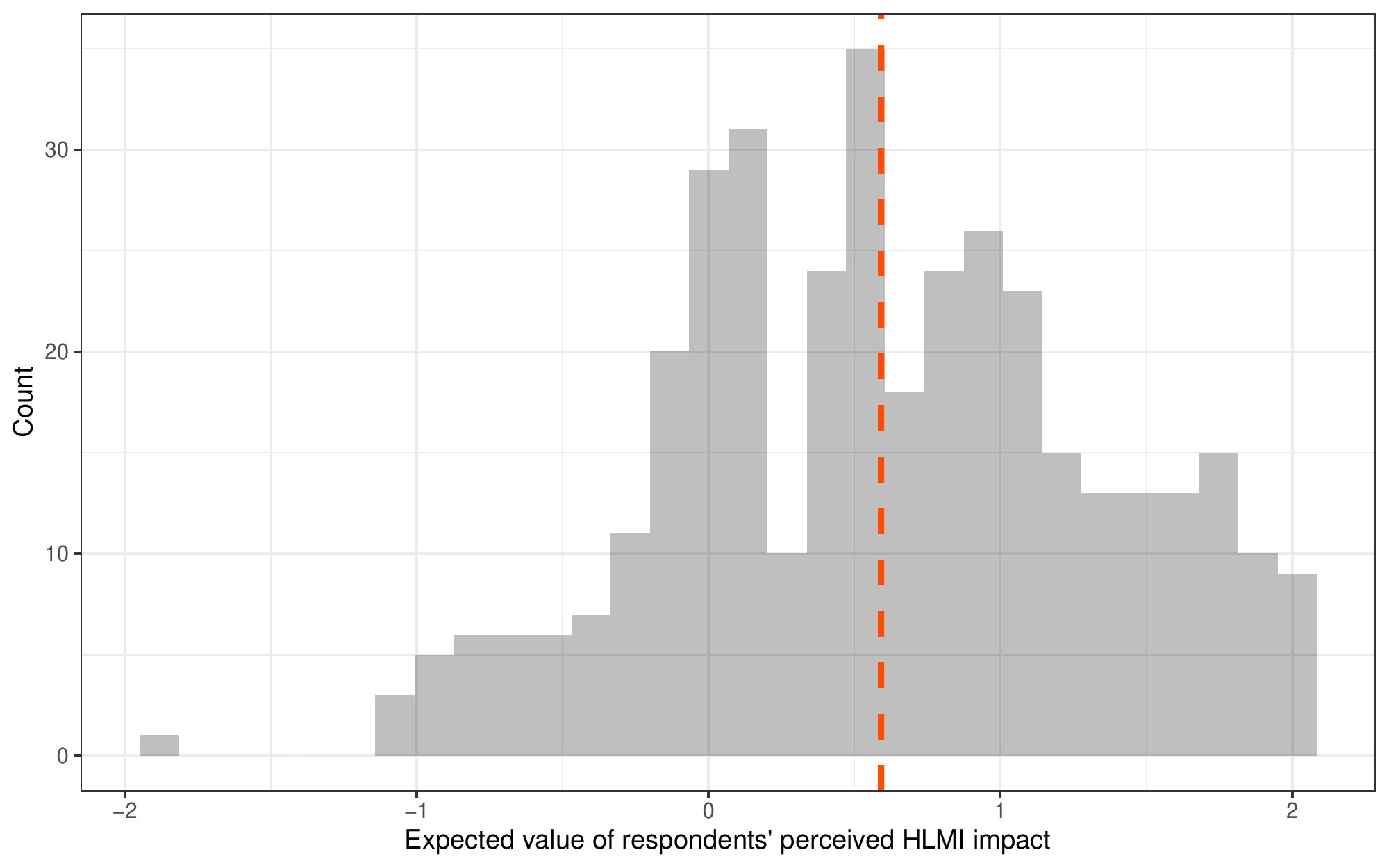}
    \caption{The distribution of the expected value for HLMI impact. The expected value is calculated by taking the sum of multiplying the value of the response (``Extremely good'' = 2 to ``Extremely bad'' = -2) by the assigned probability and dividing by 100. The dashed red line indicates the median expected value of 0.59.}
    \label{fig:hlmiimpact-expectedvalue}
\end{figure}

\FloatBarrier

\subsection{AI Progress Milestones}

\FloatBarrier

\begin{figure}[H]
    \centering
    \includegraphics[height=0.85\textheight]{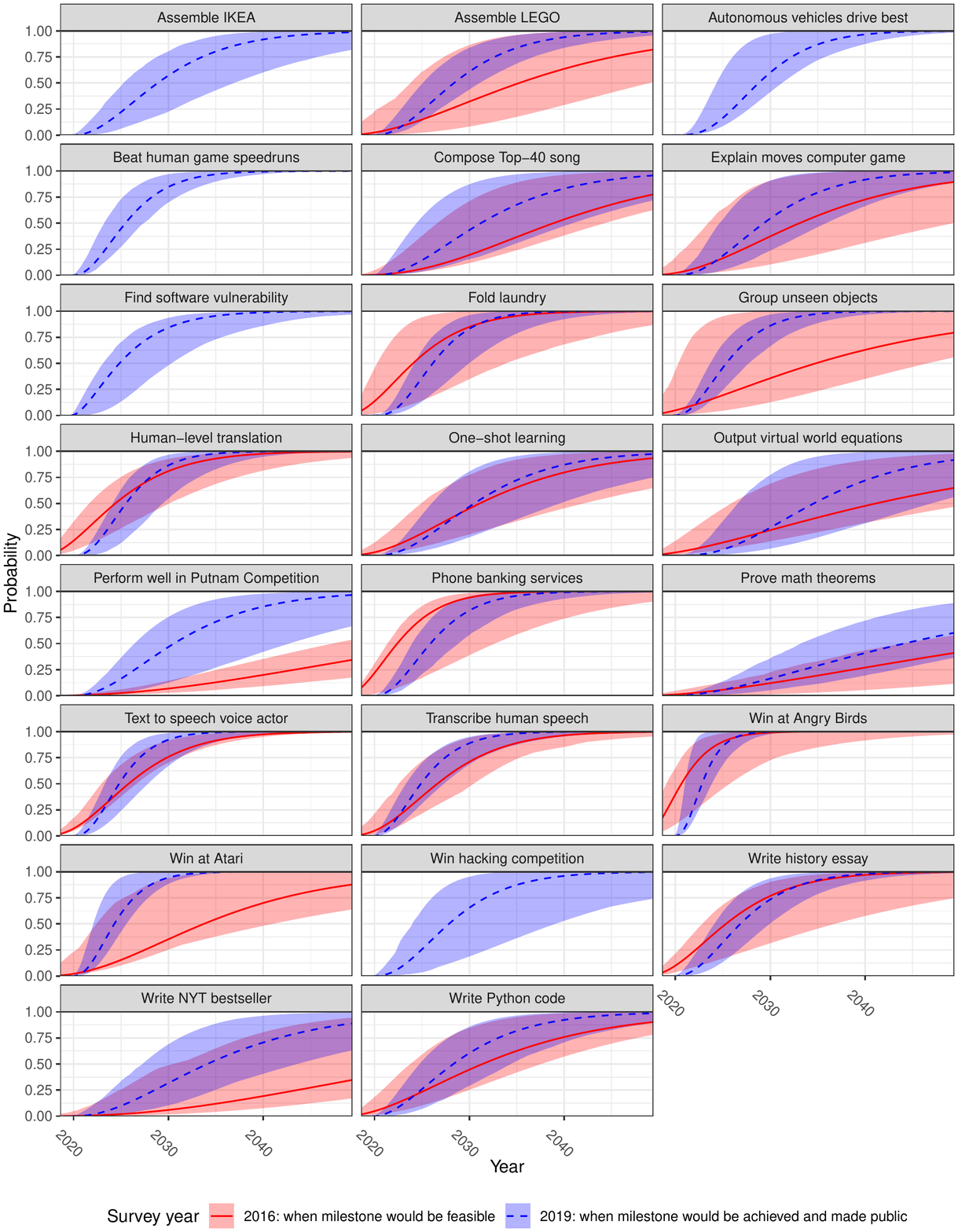}
    \caption{Aggregate forecasts of the 2016 and 2019 forecasts of AI development milestones, cross-sectional sample. The forecasts are aggregated using the median parameters method. We generated 95\% confidence intervals (CIs) for the group median CDF by bootstrapping (using 10,000 simulations) at the forecaster level.}
    \label{fig:ms_2019_nonpanel_all}
\end{figure}

\begin{figure}[H]
    \centering
    \includegraphics[width=\textwidth]{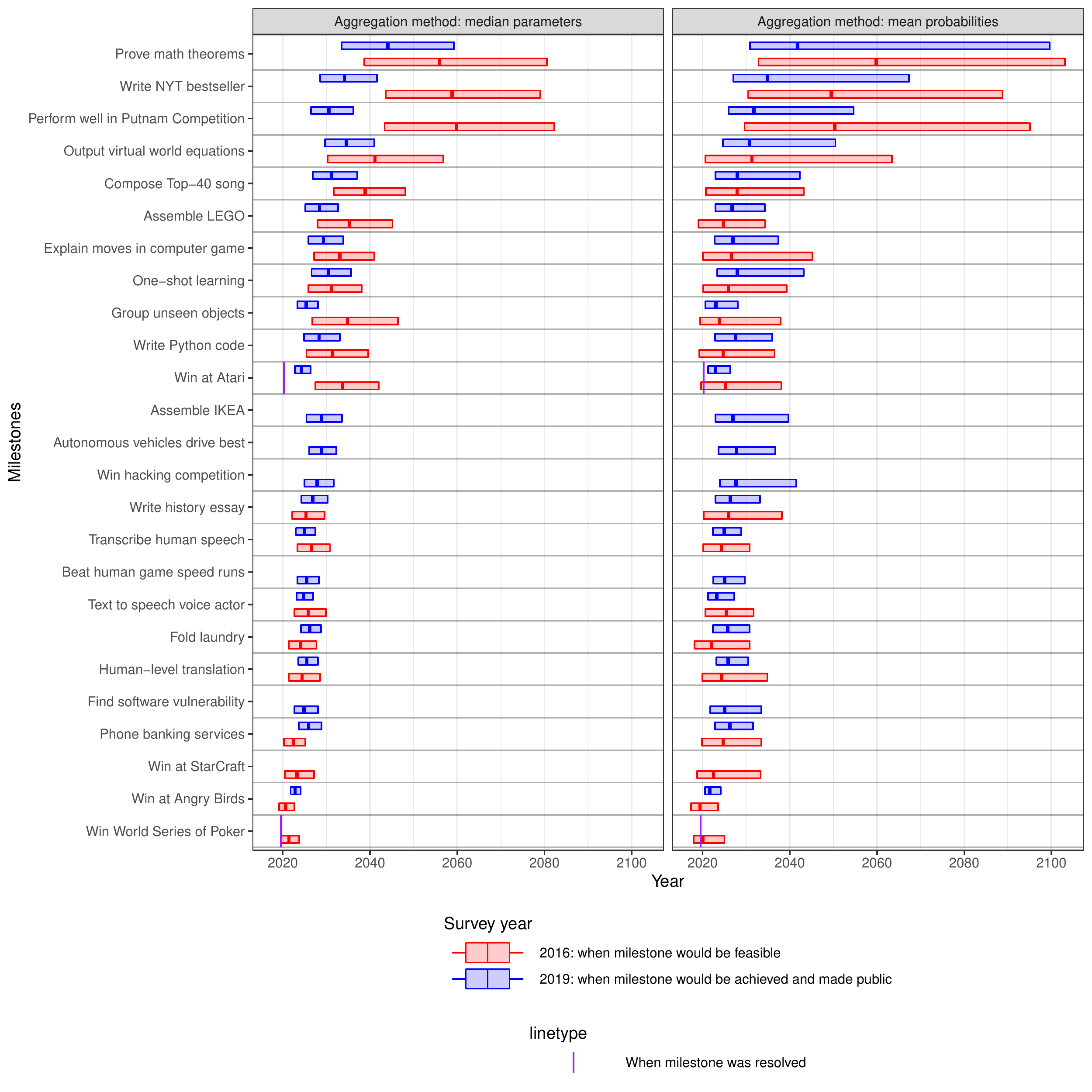}
    \caption{Comparing the 2016 and 2019 forecasts of AI development milestones, cross-sectional samples. The results in the left panel are produced using the median parameters aggregation method; the results in the right panel are produced using the mean probabilities aggregation method. Each points represents the year at 50\% probability of achieving the milestone. The lower end of the confidence interval represents 25\% probability of achieving the milestone. The upper end of the confidence interval represents 75\% probability of achieving the milestone.}
    \label{fig:my_label}
\end{figure}

\begin{figure}[H]
    \centering
    \includegraphics[width=0.9\textwidth]{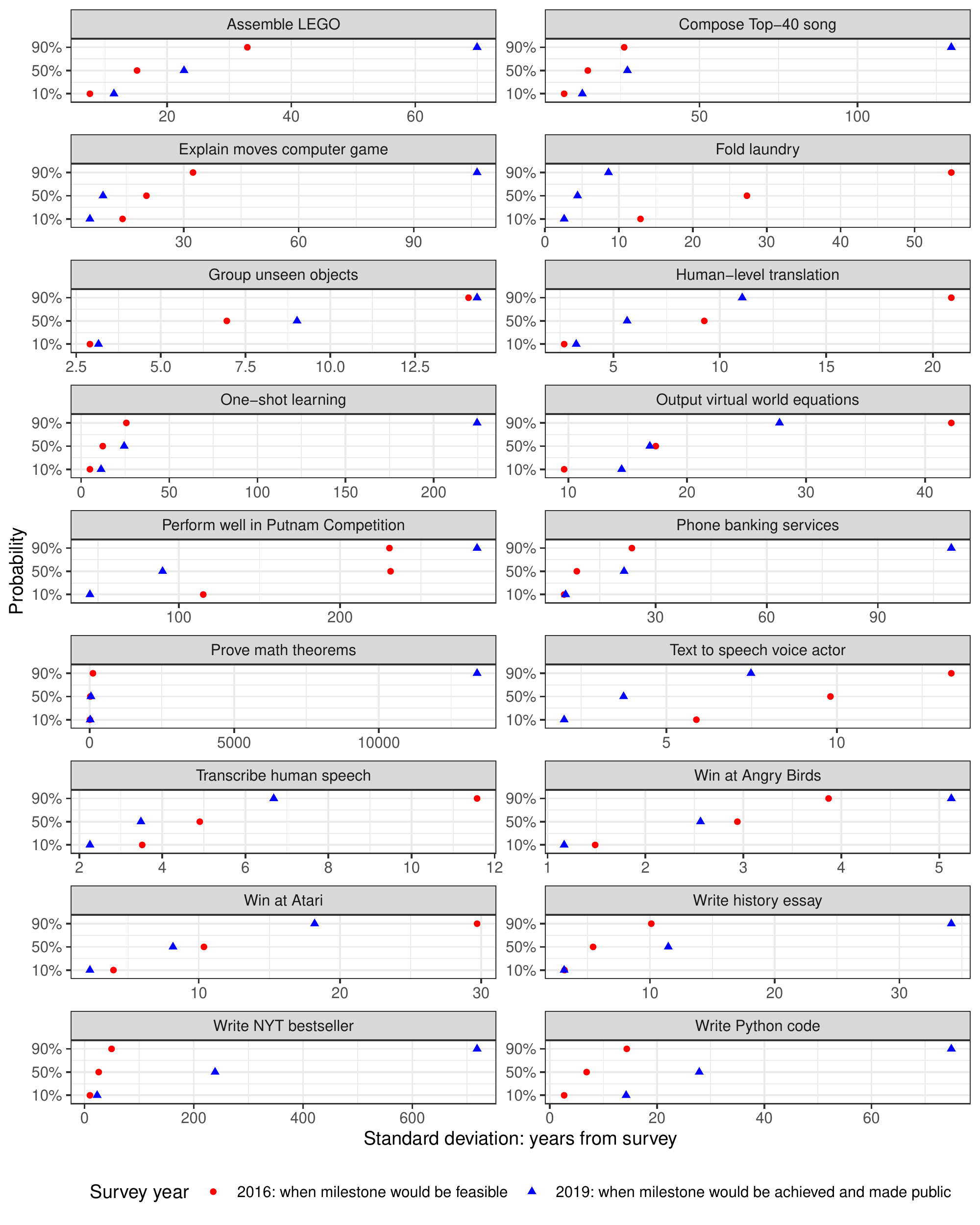}
    \caption{Comparing the dispersion of responses to the milestone questions in the 2016 and 2019 cross-sectional samples, fixed-probabilities framing. For each milestone and probability, we plot the standard deviation (years from the survey) for the 2016 sample and the 2019 sample.}
    \label{fig:std_milestones_cross}
\end{figure}

\begin{figure}[H]
    \centering
    \includegraphics[width=0.8\textwidth]{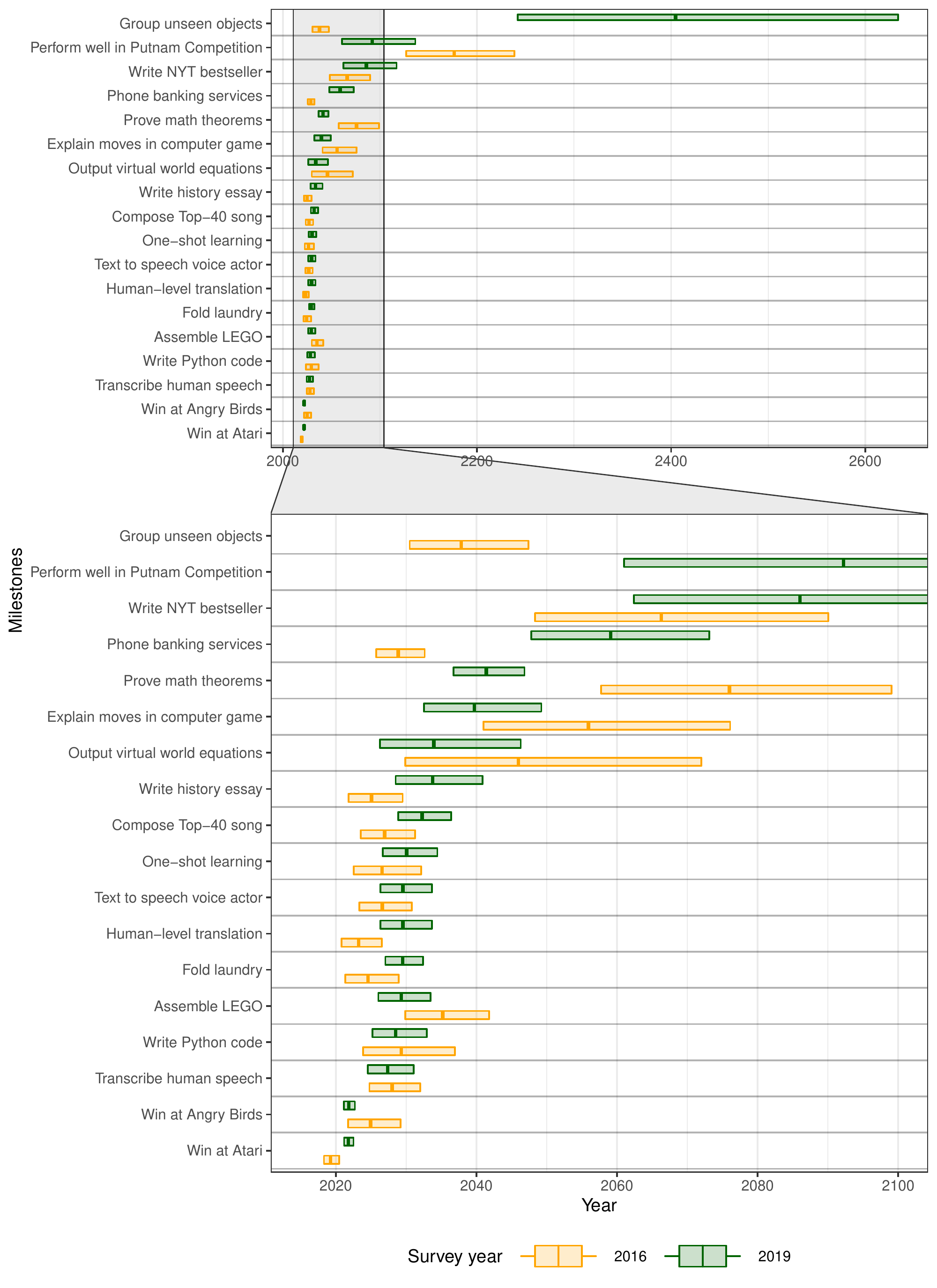}
    \caption{Comparing the 2016 and 2019 forecasts of AI development milestones, panel sample. The results are produced using the median parameters aggregation method. Each points represents the year at 50\% probability of achieving the milestone. The lower end of the confidence interval represents 25\% probability of achieving the milestone. The upper end of the confidence interval represents 75\% probability of achieving the milestone.}
    \label{fig:milestone_compare_non_panel_sample-2}
\end{figure}

\begin{figure}[H]
    \centering
    \includegraphics[width=0.8\textwidth]{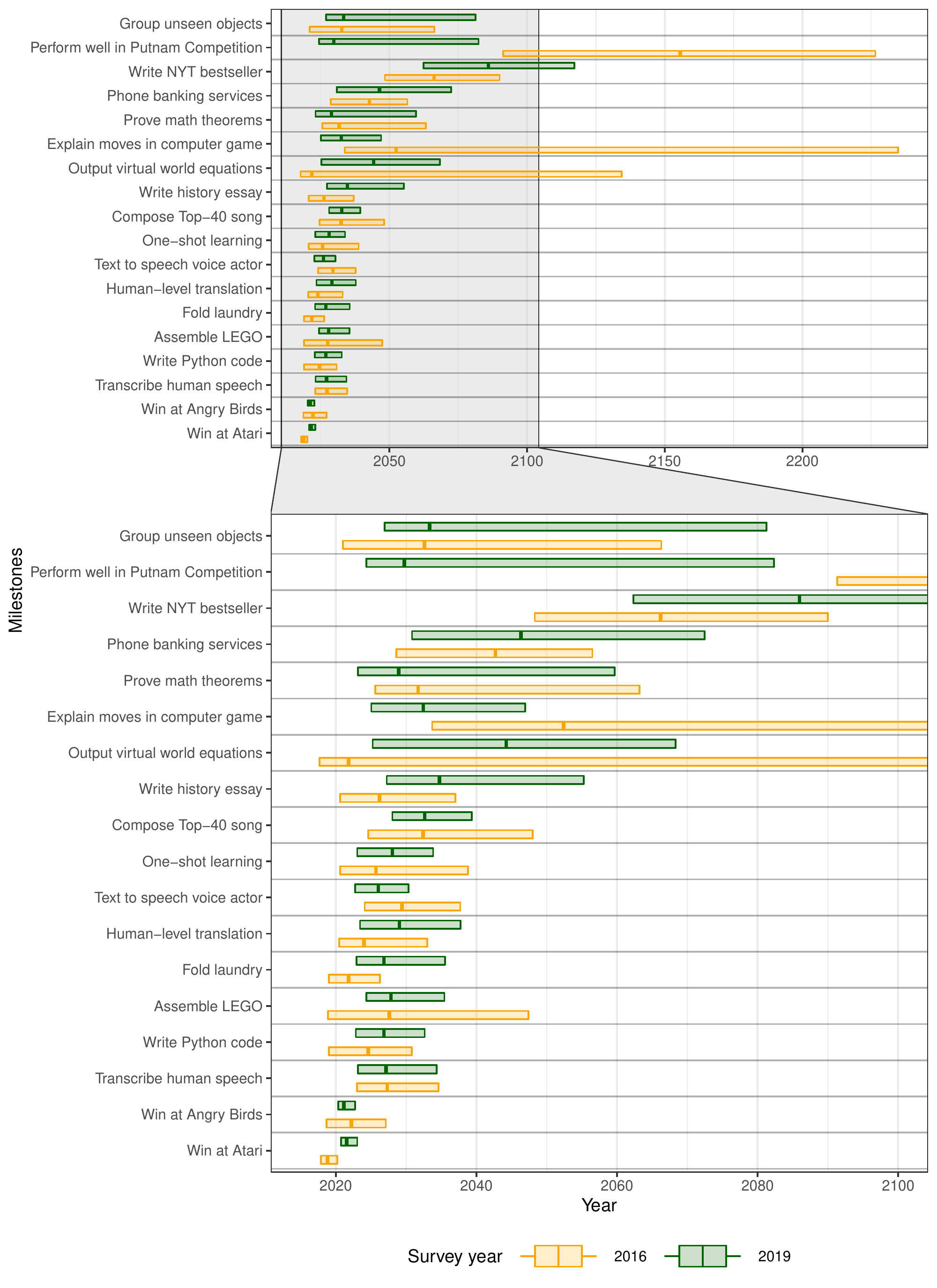}
    \caption{Comparing 2016 and 2019 forecasts of AI development milestones, panel sample. The results are produced using the mean probabilities aggregation method. Each points represents the year at 50\% probability of achieving the milestone. The lower end of the confidence interval represents 25\% probability of achieving the milestone. The upper end of the confidence interval represents 75\% probability of achieving the milestone.}
    \label{fig:milestone_compare_non_panel_sample}
\end{figure}

\begin{figure}[H]
\caption{Number of milestones resolved according to forecasts from 2016 and 2019 versus the observed number of milestones resolved. For each survey, we generated the null distribution of the number of milestones resolved according to the forecasts. Each milestone's aggregate forecast is described using a gamma distribution CDF. Whether the milestone is resolved or not is stochastic depending on the cumulative probability that the milestone is predicted to resolve by a set date. Across the milestones, we make 10000 draws using the cumulative probability that the milestone would be achieved by January 1, 2022. For each draw, we sum up the number of resolved milestones and use that as our null distribution. The one-sided \textit{p}-value is the proportion of the null distribution that is equal or greater than the observed number of resolved milestones. The two-sided \textit{p-}value is the proportion of null distribution that are at least as large as the observed number of resolved milestones in absolute value. Each histogram below show the null distribution; the dashed line show the observed number of resolved milestones. For the 2016 survey, the one-sided \textit{p}-value is 0.9547 and the two-sided \textit{p}-value is 0.9547. For the 2019 survey, the one-sided \textit{p}-value is 0.8394 and the two-sided \textit{p}-value is 0.8394.}
\label{fig:forecasts-milestone-null-dist}
\begin{subfigure}{\textwidth}
  \centering
  % include second image
  \includegraphics[width=0.75\linewidth]{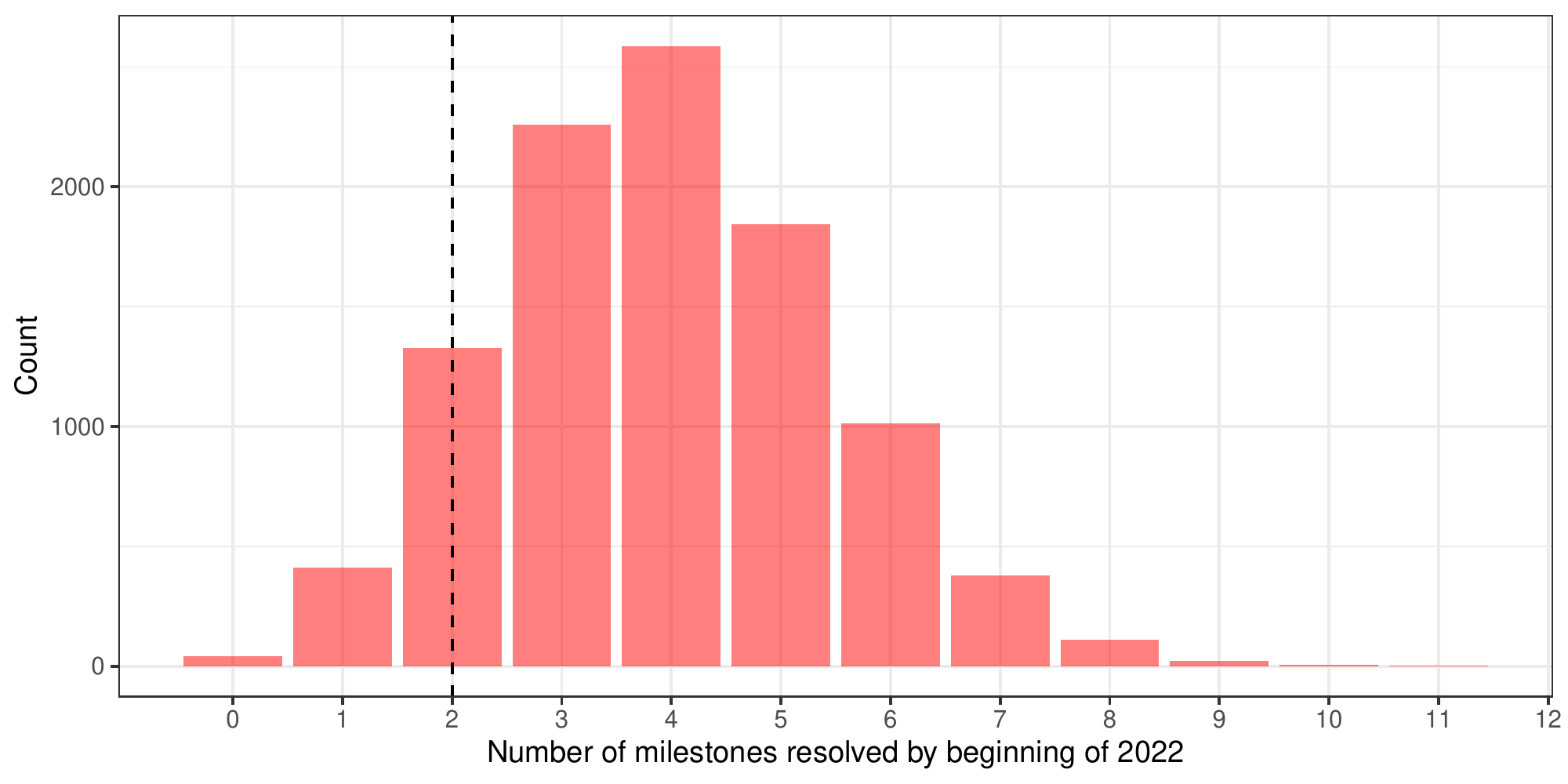}  
  \caption{2016 survey: all respondents}
  \label{fig:milestone-2016-milestones-null}
\end{subfigure}
\begin{subfigure}{\textwidth}
  \centering
  % include first image
  \includegraphics[width=0.75\linewidth]{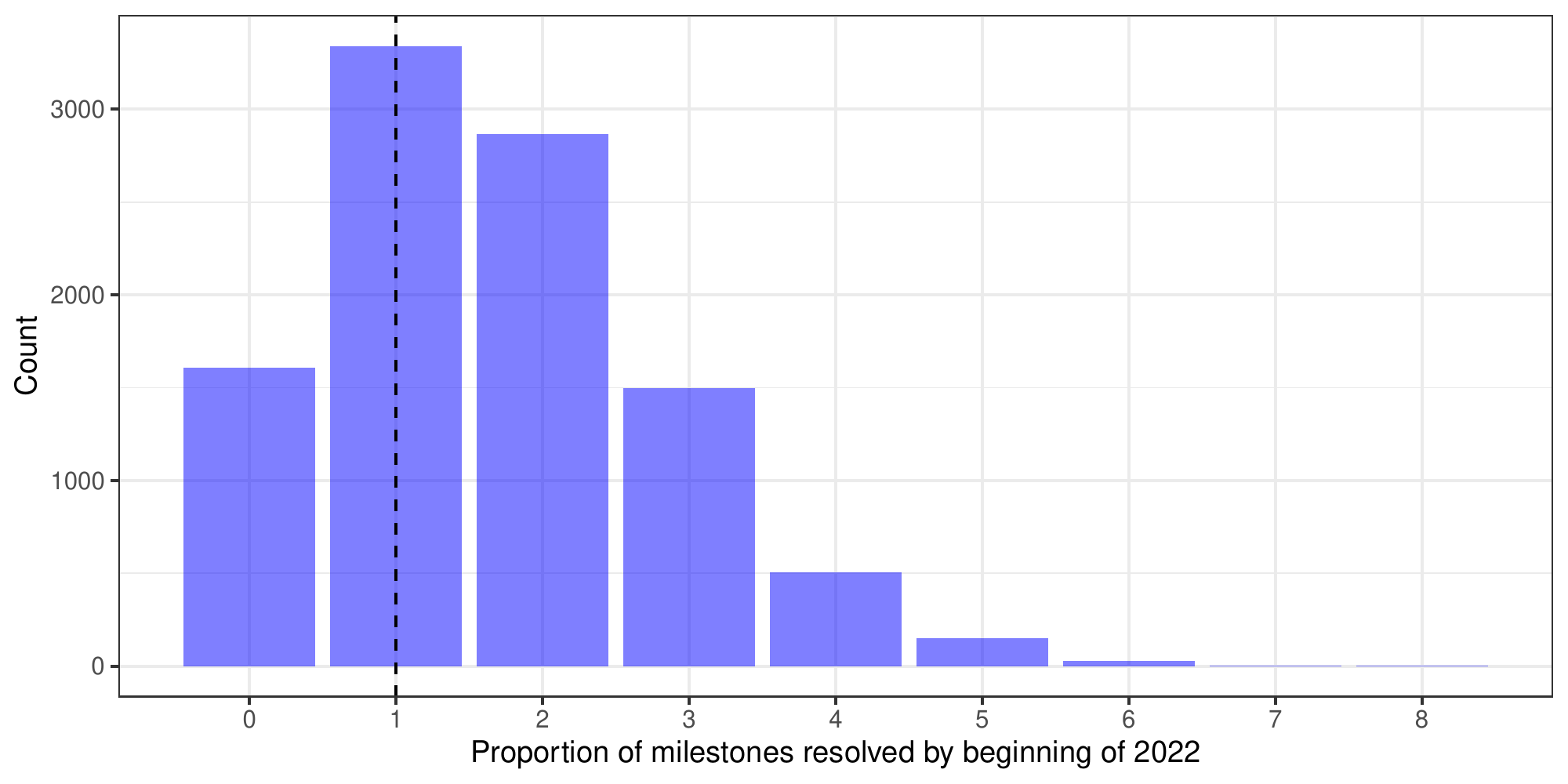}  
  \caption{2019 survey: cross-sectional sample}
  \label{fig:milestone-2019-milestones-null}
\end{subfigure}
\end{figure}

\newpage

\section{Additional Tables}

\subsection{Demographics of Survey Respondents}

\FloatBarrier

\begin{longtable}{|p{4cm}|r|r|l|p{2cm}|}
        \caption{Summary statistics of the non-respondents and respondents (all responses): binary differences. We collected demographic information for all our respondents and a random sample of 446 non-respondents using information publicly available online. The table presents the proportion of individuals in each demographic category for gender, region of undergraduate and PhD, region where the respondent works, and  the type of workplace for both non-respondents and respondents. The mean undergraduate graduation year and log citations are also shown. For each the difference between the non-respondents' and respondents' proportions is presented alongside the corresponding standard error. The Holm method was used to control the family-wise error rate. \label{tab:summarystat-cs}} \\
%\begin{tabular}{|p{4cm}|r|r|l|r|}
\hline
Variable & Non-respondent & Respondent & Difference (\textit{SE}) & Percent missing\\
\hline
Prop. male & 0.89 & 0.91 & 0.01 (0.02) & 0.01\\
\hline
Mean undergrad graduation year & 2007.62 & 2008.95 & 1.33 (0.47) & 0.21\\
\hline
Prop. undergrad region: North America & 0.25 & 0.27 & 0.02 (0.03) & 0.15\\
\hline
Prop. undergrad region: Europe & 0.26 & 0.29 & 0.02 (0.03) & 0.15\\
\hline
Prop. undergrad region: Asia & 0.43 & 0.39 & -0.04 (0.03) & 0.15\\
\hline
Prop. undergrad region: Other & 0.04 & 0.05 & 0.01 (0.01) & 0.15\\
\hline
Prop. PhD region: North America & 0.28 & 0.33 & 0.06 (0.03) & 0.08\\
\hline
Prop. PhD region: Europe & 0.59 & 0.53 & -0.07 (0.03) & 0.08\\
\hline
Prop. PhD region: Asia & 0.11 & 0.09 & -0.01 (0.02) & 0.08\\
\hline
Prop. PhD region: Other & 0.02 & 0.02 & 0.01 (0.01) & 0.08\\
\hline
Prop. currently enrolled in PhD & 0.20 & 0.33 & 0.12 (0.03)*** & 0.05\\
\hline
Mean log citations (all) & 6.75 & 6.26 & -0.49 (0.12)*** & 0.17\\
\hline
Mean h-index (all) & 19.68 & 14.42 & -5.26 (1.12)*** & 0.17\\
\hline
Prop. work region: Europe & 0.28 & 0.33 & 0.05 (0.03) & 0.01\\
\hline
Prop. work region: North America & 0.59 & 0.54 & -0.05 (0.03) & 0.01\\
\hline
Prop. work region: Asia & 0.12 & 0.12 & <0.01 (0.02) & 0.01\\
\hline
Prop. work region: Other & 0.01 & 0.02 & 0.01 (0.01) & 0.01\\
\hline
Prop. work in academia & 0.68 & 0.80 & 0.12 (0.03)*** & 0.00\\
\hline
Prop. work in industry & 0.36 & 0.35 & -0.01 (0.03) & 0.00\\
\hline
%\end{tabular}
\end{longtable}

% \clearpage

\FloatBarrier

\begin{longtable}{ll}
  \caption{Association between demographic characteristics and survey response (all contacted in 2019): results from multiple regression model. We collected demographic information for all our respondents and a random sample of 446 non-respondents using information publicly available online. Here we use multiple linear regression to predict the response to the survey using the demographic variables that we collected. The (arbitrarily chosen) reference categories, the ones that are excluded from the list of coefficients, are female/other for gender, North America for undergraduate, PhD, and work region, and industry for type of workplace. The \textit{F}-test of overall significance rejects the null hypothesis that respondents do not differ in whether they responded to the survey depending on demographic characteristics. The Holm method was used to control the family-wise error rate.} 
  \label{tab:cs-response-regression-analysis} \\
\hline
  & Coefficient (\textit{SE}) \\
\hline
(Intercept) & 0.540$^{***}$ \\ 
  & (0.016) \\ 
  Male & 0.022 \\ 
  & (0.016) \\ 
  Undergrad graduation year & 0.002 \\ 
  & (0.021) \\ 
  Undergrad region: Europe & $-$0.030 \\ 
  & (0.022) \\ 
  Undergrad region: Asia & $-$0.037 \\ 
  & (0.020) \\ 
  Undergrad region: Other & $-$0.013 \\ 
  & (0.018) \\ 
  PhD region: Europe & 0.039 \\ 
  & (0.026) \\ 
  PhD region: Asia & $-$0.019 \\ 
  & (0.023) \\ 
  PhD region: Other & 0.009 \\ 
  & (0.021) \\ 
  Currently enrolled in PhD & 0.033 \\ 
  & (0.019) \\ 
  Log all citations & 0.022 \\ 
  & (0.028) \\ 
  All h-index & $-$0.083$^{*}$ \\ 
  & (0.026) \\ 
  Work region: Europe & 0.017 \\ 
  & (0.026) \\ 
  Work region: Asia & 0.034 \\ 
  & (0.024) \\ 
  Work region: Other & 0.014 \\ 
  & (0.021) \\ 
  Work in academia & 0.069$^{***}$ \\ 
  & (0.017) \\ 
  Missing: undergrad year & 0.001 \\ 
  & (0.028) \\ 
  Missing: undergrad region & $-$0.044 \\ 
  & (0.028) \\ 
  Missing: all citations & 0.009 \\ 
  & (0.016) \\ 
\hline
Number of respondents & 970\\
\textit{F}-statistic & 4.196$^{***}$ (df = 18, 951)  \\
\hline \\[-1.8ex] 
\multicolumn{2}{l}{$^{*}$p $<$ .05; $^{**}$p $<$ .01; $^{***}$p $<$ .001} \\ 
\end{longtable} 

\begin{table}[H]
\caption{Summary statistics of the non-respondents and respondents (all contacted in 2019): binary differences. We used demographic data collected for \cite{GraceEtAl} for this analysis. The table presents the proportion of individuals in each demographic category for gender, citation count in 2016, PhD start year, country of undergraduate degree, and region of undergraduate degree. The mean undergraduate graduation year and log citations are also shown. For each the difference between the non-respondents' and respondents' proportions is presented alongside the corresponding standard error. The Holm method was used to control the family-wise error rate. \label{tab:summarystat-panel-sample}}
\begin{tabular}{|p{4cm}|r|r|l|p{2cm}|}
\hline
Variable & Non-respondent & Respondent & Difference (\textit{SE}) & Proportion missing\\
\hline
Prop. male & 0.95 & 0.94 & -0.01 (0.03) & 0.16\\
\hline
Citation count in 2016 & 2728.91 & 2687.47 & -41.44 (975.57) & 0.16\\
\hline
PhD start year & 2007.54 & 2006.59 & -0.95 (0.83) & 0.25\\
\hline
Prop. undergrad country: the US & 0.15 & 0.28 & 0.13 (0.05) & 0.21\\
\hline
Prop. undergrad country: China & 0.15 & 0.08 & -0.07 (0.04) & 0.21\\
\hline
Prop. undergrad country: Other & 0.49 & 0.43 & -0.05 (0.06) & 0.21\\
\hline
Prop. undergrad region: Europe & 0.22 & 0.27 & 0.05 (0.05) & 0.21\\
\hline
Prop. undergrad region: Asia & 0.29 & 0.14 & -0.16 (0.05)** & 0.21\\
\hline
Prop. undergrad region: North America & 0.20 & 0.30 & 0.09 (0.05) & 0.21\\
\hline
Prop. undergrad region: Other & 0.07 & 0.09 & 0.02 (0.03) & 0.21\\
\hline
\end{tabular}
\end{table}

\begin{table}[H] \centering 
  \caption{Association between demographic characteristics and survey response (panel sample): results from multiple regression model. We used demographic data collected for \cite{GraceEtAl} to predict which researchers who completed some part of the 2016 completed some part of the 2019 survey. The (arbitrarily chosen) reference categories, the ones that are excluded from the list of coefficients, are female/other for gender; countries that are not US or China for undergraduate country; regions that are not Europe, Asia, or North America for undergraduate region. The \textit{F}-test of overall significance rejects the null hypothesis that respondents do not differ in whether they responded to the survey depending on demographic characteristics. The Holm method was used to control the family-wise error rate.} 
  \label{tab:panel-who-answered} 
\begin{tabular}{@{\extracolsep{5pt}}lc} 
\\[-1.8ex]\hline \\[-1.8ex] 
\\[-1.8ex] & Coefficient (\textit{SE}) \\ 
\hline \\[-1.8ex] 
 (Intercept) & 14.319 \\ 
  & (8.559) \\ 
  Male & $-$0.018 \\ 
  & (0.086) \\ 
  Log(citation count in 2016) & $-$0.043 \\ 
  & (0.025) \\ 
  PhD start year & $-$0.007 \\ 
  & (0.004) \\ 
  Undergraduate country: US & 0.199 \\ 
  & (0.080) \\ 
  Undergraduate country: China & 0.018 \\ 
  & (0.057) \\ 
  Undergraduate country: missing & $-$0.103 \\ 
  & (0.089) \\ 
  Undergraduate region: Europe & $-$0.009 \\ 
  & (0.082) \\ 
  Undergraduate region: Asia & $-$0.131 \\ 
  & (0.083) \\ 
  Undergraduate region: North America & $-$0.131 \\ 
  & (0.096) \\ 
  Missing: gender & $-$0.003 \\ 
  & (0.086) \\ 
  Missing: citation count in 2016 & 0.078 \\ 
  & (0.092) \\ 
  Missing: PhD start year & $-$0.053 \\ 
  & (0.050) \\ 
  \hline \\[-1.8ex] 
 Number of respondents & 472 \\ 
\textit{F}-statistic & 1.981$^{*}$ (df = 12; 459) \\ 
\hline \\[-1.8ex] 
\multicolumn{2}{l}{$^{*}$p $<$ .05; $^{**}$p $<$ .01; $^{***}$p $<$ .001} \\ 
\end{tabular} 
\end{table}

\subsection{Human/high-level machine intelligence forecasts}

\begin{table}[H]

\caption{Predicted year of human/high-level machine intelligence by year, by different samples}
\centering
\begin{tabular}[t]{|r|p{3cm}|p{3cm}|p{3cm}|p{3cm}|}
\hline
Probaility & Cross-sectional sample 2019 & Cross-sectional sample 2016 & Panel sample 2019 & Panel sample 2016\\
\hline
0.1 & 2034 & 2032 & 2042 & 2037\\
\hline
0.2 & 2041 & 2039 & 2051 & 2044\\
\hline
0.3 & 2047 & 2045 & 2060 & 2050\\
\hline
0.4 & 2054 & 2051 & 2068 & 2056\\
\hline
0.5 & 2060 & 2058 & 2076 & 2062\\
\hline
0.6 & 2068 & 2065 & 2085 & 2069\\
\hline
0.7 & 2076 & 2074 & 2097 & 2077\\
\hline
0.8 & 2088 & 2086 & 2111 & 2088\\
\hline
0.9 & 2107 & 2106 & 2134 & 2105\\
\hline
\end{tabular}
\end{table}

\begin{table}[H]

\caption{Predicted probability of human/high-level machine intelligence by year, by different samples}
\centering
\begin{tabular}[t]{|r|p{3cm}|p{3cm}|p{3cm}|p{3cm}|}
\hline
Year & Cross-sectional sample 2019 & Cross-sectional sample 2016 & Panel sample 2019 & Panel sample 2016\\
\hline
2025 & 0.012 & 0.020 & 0.003 & 0.006\\
\hline
2030 & 0.050 & 0.071 & 0.016 & 0.032\\
\hline
2035 & 0.108 & 0.141 & 0.042 & 0.080\\
\hline
2040 & 0.181 & 0.220 & 0.079 & 0.146\\
\hline
2045 & 0.261 & 0.303 & 0.127 & 0.224\\
\hline
2050 & 0.342 & 0.384 & 0.183 & 0.307\\
\hline
2055 & 0.422 & 0.462 & 0.242 & 0.390\\
\hline
2060 & 0.497 & 0.534 & 0.305 & 0.470\\
\hline
2065 & 0.566 & 0.598 & 0.368 & 0.546\\
\hline
2070 & 0.629 & 0.656 & 0.429 & 0.614\\
\hline
2075 & 0.685 & 0.708 & 0.488 & 0.675\\
\hline
2080 & 0.734 & 0.752 & 0.544 & 0.728\\
\hline
\end{tabular}
\end{table}

\begin{table}[H]
\caption{Summary results: panel sample, fixed probabilities framing}
\begin{tabular}[t]{|p{2.5cm}|p{2.5cm}|p{2.5cm}|p{2.5cm}|p{2.5cm}|r|}
\hline
Probability & Mean year 2016 & Median year 2016 & Mean year 2019 & Median year 2019 & $n$\\
\hline
0.1 & 75.26 & 17 & 42.22 & 20 & 27\\
\hline
0.5 & 453.11 & 47 & 125.19 & 40 & 27\\
\hline
0.9 & 11314.78 & 97 & 625.33 & 100 & 27\\
\hline
\end{tabular}
\end{table}

\begin{table}[H]
\caption{Summary results: panel sample, fixed years framing}
\begin{tabular}[t]{|p{2.5cm}|p{2.5cm}|p{2.5cm}|p{2.5cm}|p{2.5cm}|r|}
\hline
Year from survey year & Mean prob 2016 & Median prob 2016 & Mean prob 2019 & Median prob 2019 & $n$\\
\hline
10 & 0.09 & 0.02 & 0.05 & 0.01 & 22\\
\hline
20 & 0.21 & 0.10 & 0.11 & 0.04 & 22\\
\hline
40 & 0.42 & 0.22 & 0.25 & 0.10 & 22\\
\hline
\end{tabular}
\end{table}

\subsection{Changes in HLMI forecasts: 2016 vs. 2019}

\begin{table}[H]

\caption{Comparing all 2016 respondents' median parameters aggregate CDF with all 2019 respondents' median parameters aggregate CDF}
\begin{tabular}[t]{|p{5cm}|r|l|l|}
\hline
Outcome & Results & $p$-value & Holm correction (p < 0.05)\\
\hline
Wasserstein metric observed value & 2.913 &  & \\
\hline
Wasserstein metric, p-value (1-tailed) & 0.289 & 0.289 & No\\
\hline
Wasserstein metric, p-value (2-tailed) & 0.289 & 0.289 & No\\
\hline
Kolmogorov-Smirnov statistic, 2-sided, observed value & 0.068 &  & \\
\hline
Kolmogorov-Smirnov statistic, 2-sided, p-value (1-tailed) & 0.329 & 0.329 & No\\
\hline
Kolmogorov-Smirnov statistic, 2-sided, p-value (2-tailed) & 0.329 & 0.329 & No\\
\hline
Kolmogorov-Smirnov statistic, 1-sided, observed value & 0.068 &  & \\
\hline
Kolmogorov-Smirnov statistic, 1-sided, p-value (1-tailed) & 0.180 & 0.180 & No\\
\hline
Kolmogorov-Smirnov statistic, 1-sided, p-value (2-tailed) & 0.180 & 0.180 & No\\
\hline
\end{tabular}
\end{table}

\begin{table}[H]

\caption{Comparing all 2016 respondents' median parameters aggregate CDF with 2019 definition (human-level machine intelligence) respondents' median parameters aggregate CDF}
\centering
\begin{tabular}[t]{|p{5cm}|r|r|l|}
\hline
Outcome & Results & p-value & Holm correction (p < 0.05)\\
\hline
Wasserstein metric observed value & 1.766 &  & \\
\hline
Wasserstein metric, p-value (1-tailed) & 0.597 & 0.597 & No\\
\hline
Wasserstein metric, p-value (2-tailed) & 0.597 & 0.597 & No\\
\hline
Kolmogorov-Smirnov statistic, 2-sided, observed value & 0.043 &  & \\
\hline
Kolmogorov-Smirnov statistic, 2-sided, p-value (1-tailed) & 0.648 & 0.648 & No\\
\hline
Kolmogorov-Smirnov statistic, 2-sided, p-value (2-tailed) & 0.648 & 0.648 & No\\
\hline
Kolmogorov-Smirnov statistic, 1-sided, observed value & 0.043 &  & \\
\hline
Kolmogorov-Smirnov statistic, 1-sided, p-value (1-tailed) & 0.345 & 0.345 & No\\
\hline
Kolmogorov-Smirnov statistic, 1-sided, p-value (2-tailed) & 0.346 & 0.346 & No\\
\hline
\end{tabular}
\end{table}

\begin{table}[H]

\caption{Comparing panel 2016 respondents' median parameters aggregate CDF with panel 2019 respondents' median parameters aggregate CDF}
\centering
\begin{tabular}[t]{|p{5cm}|r|r|l|}
\hline
Outcome & Results & p-value & Holm correction (p < 0.05)\\
\hline
Wasserstein metric observed value & 6.460 &  & \\
\hline
Wasserstein metric, p-value (1-tailed) & 0.101 & 0.101 & No\\
\hline
Wasserstein metric, p-value (2-tailed) & 0.101 & 0.101 & No\\
\hline
Kolmogorov-Smirnov statistic, 2-sided, observed value & 0.187 &  & \\
\hline
Kolmogorov-Smirnov statistic, 2-sided, p-value (1-tailed) & 0.111 & 0.111 & No\\
\hline
Kolmogorov-Smirnov statistic, 2-sided, p-value (2-tailed) & 0.111 & 0.111 & No\\
\hline
Kolmogorov-Smirnov statistic, 1-sided, observed value & 0.187 &  & \\
\hline
Kolmogorov-Smirnov statistic, 1-sided, p-value (1-tailed) & 0.061 & 0.061 & No\\
\hline
Kolmogorov-Smirnov statistic, 1-sided, p-value (2-tailed) & 0.061 & 0.061 & No\\
\hline
\end{tabular}
\end{table}

\subsection{Changes in HLMI forecasts and AI progress milestone process in Panel Sample: 2016 vs. 2019}

\begin{table}[H]
\caption{Comparing the panel sample results by framing type through clustered Wilcoxon signed rank tests using Rosner-Glynn-Lee method. For the HLMI forecasts, each cluster is the respondent. For the milestones forecasts, each cluster is the respondent-milestone. For the milestone forecasts, each respondent is randomly assigned three milestones to forecast. \label{tab:wilcoxon-results}}
\begin{tabular}{|p{2cm}|l|p{2cm}|l|l|l|}
\hline
Framing type      & \textit{Z}        & \textit{p}-value (two-sided) & Pairs of observations & Clusters & Forecast type \\ \hline
Fixed probabilities & -0.17616 & 0.8602  & 81                    & 27 & HLMI          \\ \hline
Fixed years        & 2.2304   & 0.02572 & 66                    & 22 & HLMI          \\ \hline
Fixed probabilities & -2.0383 & 0.04152 & 174 & 58 & Milestones \\
\hline 
\end{tabular}
\end{table}

%\begin{figure}[ht]
%    \centering
%    \includegraphics[height=0.9\textheight]{images/all2019ms-curves-1.pdf}
%    \caption{Milestone curves 2019}
%    \label{fig:all_curves}
%\end{figure}

% \begin{figure}
%     \centering
%     \includegraphics[width=180mm]{images/hlmiimpact-goodbad.png}
%     \caption{Combined  probabilities of both extremely and on balance good, and combined probabilities of extremely and on balance bad. The figures shows the histogram of responses}
%     \label{fig:hlmiimpactgoodbadhist}
% \end{figure}

\subsubsection{Impact of HLMI}

\begin{table}[H]

\caption{Predicting expected value of perceived impact based on definition of HLMI ('high-level' vs. 'human-level')}
\centering
\begin{tabular}[t]{lc}
\toprule
  & Coefficient (\textit{SE})\\
\midrule
(Intercept) & 0.632***\\
 & (0.042)\\
'High-level' definition & -0.152\\
 & (0.102)\\
\midrule
Number of respondents & 373\\
$F$-statistic & 2.229 on 1 and 371 DF,  $p$-value: 0.1363 \\
\bottomrule
\multicolumn{2}{l}{\textsuperscript{} * p < 0.05, ** p < 0.01, *** p < 0.01}\\
\end{tabular}
\end{table}

\begin{table}[H]

\caption{Comparing median percentages on the impact of HLMI question: 2016 survey versus 2019 survey}
\centering
\begin{tabular}[t]{|l|p{2cm}|p{2cm}|p{2cm}|p{2cm}|p{2cm}|}
\hline
Year & Extremely good (2) & On balance good (1) & More or less neutral (0) & On  balance bad (-1) & Extremely bad (-2) \\
\hline
2016 & 20 & 25 & 20 & 10 & 5\\
\hline
2019 & 20 & 27 & 16 & 10 & 2\\
\hline
\end{tabular}
\end{table}

\begin{table}[H]

\caption{Difference in expected value of HLMI's societal impact}
\centering
\begin{tabular}[t]{lp{5cm}p{5cm}}
\toprule
  & Model 1: all responses & Model 2: panel sample\\
\midrule
(Intercept) & 0.515*** & 0.478***\\
 & (0.038) & (0.086)\\
Survey year: 2019 & 0.090 & -0.006\\
 & (0.051) & (0.092)\\
\midrule
Number of observations & 727 & 136\\
Number of unique respondents & 659 & 68 \\
$F$-statistic & 3.111 on 1 and 658 DF,  $p$-value: 0.07825 & 0.004846 on 1 and 67 DF,  $p$-value: 0.9447 \\
\bottomrule
\multicolumn{3}{l}{\textsuperscript{} * p < 0.05, ** p < 0.01, *** p < 0.01}\\
\end{tabular}
\end{table}

\subsection{AI Progress Milestones}

\begin{longtable}{|p{4cm}|p{2cm}|p{1cm}|p{1cm}|p{1cm}|p{1.4cm}|p{1cm}|p{1cm}|r|}

\caption{Distribution of AI development milestones forecasts, 2019 survey, cross-sectional sample, fixed probabilities framing} \\

\hline
Milestones & Probability & Min year & Q1 year & Mean year & Median year & Q3 year & Max year & $n$\\
\hline
Assemble IKEA & 0.1 & 1.0 & 2.0 & 3.3 & 3.0 & 4.2 & 10 & 20\\
\hline
Assemble IKEA & 0.5 & 2.0 & 4.8 & 7.1 & 5.5 & 8.0 & 30 & 20\\
\hline
Assemble IKEA & 0.9 & 3.0 & 8.0 & 15.2 & 10.0 & 20.0 & 50 & 20\\
\hline
Assemble LEGO & 0.1 & 1.0 & 2.0 & 6.1 & 4.0 & 5.0 & 50 & 17\\
\hline
Assemble LEGO & 0.5 & 2.0 & 4.0 & 12.9 & 9.0 & 10.0 & 100 & 17\\
\hline
Assemble LEGO & 0.9 & 3.0 & 7.0 & 32.6 & 14.0 & 20.0 & 300 & 17\\
\hline

\end{longtable}

\begin{longtable}{|p{4cm}|p{2cm}|p{1cm}|p{1cm}|p{1cm}|p{1.4cm}|p{1cm}|p{1cm}|r|}

\caption{Distribution of AI development milestones forecasts, 2019 survey, cross-sectional sample, fixed years framing} \\

\hline
Milestones & Year from survey year & Min prob & Q1 prob & Mean prob & Median prob & Q3 prob & Max prob & $n$\\
\hline
Assemble IKEA & 1 & 0.00 & 0.01 & 0.15 & 0.05 & 0.20 & 1.00 & 25\\
\hline
Assemble IKEA & 5 & 0.00 & 0.25 & 0.50 & 0.50 & 0.80 & 1.00 & 25\\
\hline
Assemble IKEA & 10 & 0.00 & 0.50 & 0.70 & 0.90 & 0.99 & 1.00 & 25\\
\hline
Assemble LEGO & 1 & 0.00 & 0.01 & 0.11 & 0.10 & 0.20 & 0.50 & 17\\
\hline
Assemble LEGO & 5 & 0.01 & 0.05 & 0.38 & 0.30 & 0.60 & 0.90 & 17\\
\hline
Assemble LEGO & 10 & 0.05 & 0.20 & 0.59 & 0.60 & 0.95 & 0.99 & 17\\
\hline
\end{longtable}

\begin{longtable}{|p{4cm}|p{2cm}|p{1cm}|p{1cm}|p{1cm}|p{1.4cm}|p{1cm}|p{1cm}|r|}

\caption{Distribution of AI development milestones forecasts, 2016 survey, cross-sectional sample, fixed probabilities framing} \\

\hline
Milestones & Probability & Min year & Q1 year & Mean year & Median year & Q3 year & Max year & \textit{n}\\
\hline
Assemble LEGO & 0.1 & 0 & 3.0 & 6.2 & 5.0 & 5.0 & 30 & 13\\
\hline
Assemble LEGO & 0.5 & 0 & 5.0 & 13.5 & 10.0 & 15.0 & 60 & 13\\
\hline
Assemble LEGO & 0.9 & 2 & 10.0 & 29.2 & 15.0 & 30.0 & 100 & 13\\
\hline
Compose Top-40 song & 0.1 & 1 & 3.0 & 7.5 & 5.0 & 10.0 & 30 & 20\\
\hline
Compose Top-40 song & 0.5 & 2 & 5.8 & 15.8 & 10.0 & 20.0 & 50 & 20\\
\hline
Compose Top-40 song & 0.9 & 3 & 11.0 & 28.8 & 20.0 & 50.0 & 100 & 20\\
\hline

\end{longtable}

\begin{longtable}{|p{4cm}|p{2cm}|p{1cm}|p{1cm}|p{1cm}|p{1.4cm}|p{1cm}|p{1cm}|r|}

\caption{Distribution of AI development milestones forecasts, 2016 survey, cross-sectional sample, fixed years framing} \\

\hline
Milestones & Year from 2019 & Min prob & Q1 prob & Mean prob & Median prob & Q3 prob & Max prob & \textit{n} \\
\hline
Assemble LEGO & 10 & 0.05 & 0.30 & 0.56 & 0.58 & 0.89 & 1.00 & 22\\
\hline
Assemble LEGO & 20 & 0.20 & 0.50 & 0.74 & 0.85 & 0.95 & 1.00 & 22\\
\hline
Assemble LEGO & 50 & 0.40 & 0.95 & 0.92 & 0.99 & 1.00 & 1.00 & 22\\
\hline
Compose Top-40 song & 10 & 0.00 & 0.12 & 0.42 & 0.28 & 0.78 & 1.00 & 18\\
\hline
Compose Top-40 song & 20 & 0.10 & 0.43 & 0.61 & 0.50 & 0.89 & 1.00 & 18\\
\hline
Compose Top-40 song & 50 & 0.15 & 0.90 & 0.87 & 0.90 & 0.98 & 1.00 & 18\\
\hline

\end{longtable}

\begin{longtable}{|p{4cm}|p{2cm}|p{1cm}|p{1cm}|p{1cm}|p{1.4cm}|p{1cm}|p{1cm}|r|}

\caption{Distribution of AI development milestones forecasts, 2019 survey, panel sample, fixed probabilities framing} \\

\hline
Milestones & Probability & Min year & Q1 year & Mean year & Median year & Q3 year & Max year & $n$\\
\hline
Assemble LEGO & 0.1 & 2.0 & 2.0 & 4.8 & 3.5 & 6.2 & 10 & 4\\
\hline
Assemble LEGO & 0.5 & 5.0 & 5.0 & 11.2 & 7.5 & 13.8 & 25 & 4\\
\hline
Assemble LEGO & 0.9 & 8.0 & 13.2 & 23.2 & 17.5 & 27.5 & 50 & 4\\
\hline
Compose Top-40 song & 0.1 & 5.0 & 5.0 & 7.0 & 6.5 & 8.5 & 10 & 4\\
\hline
Compose Top-40 song & 0.5 & 7.0 & 9.2 & 11.8 & 12.5 & 15.0 & 15 & 4\\
\hline
Compose Top-40 song & 0.9 & 12.0 & 16.5 & 37.2 & 19.0 & 39.8 & 99 & 4\\
\hline
\end{longtable}

\begin{longtable}{|p{4cm}|p{2cm}|p{1cm}|p{1cm}|p{1cm}|p{1.4cm}|p{1cm}|p{1cm}|r|}

\caption{Distribution of AI development milestones forecasts, 2019 survey, panel sample, fixed years framing} \\

\hline
Milestones & Year from survey year & Min prob & Q1 prob & Mean prob & Median prob & Q3 prob & Max prob & $n$\\
\hline
Compose Top-40 song & 1 & 0.00 & 0.00 & 0.00 & 0.00 & 0.00 & 0.00 & 1\\
\hline
Compose Top-40 song & 5 & 0.05 & 0.05 & 0.05 & 0.05 & 0.05 & 0.05 & 1\\
\hline
Compose Top-40 song & 10 & 0.10 & 0.10 & 0.10 & 0.10 & 0.10 & 0.10 & 1\\
\hline
Explain moves in computer game & 1 & 0.00 & 0.03 & 0.12 & 0.05 & 0.17 & 0.30 & 3\\
\hline
Explain moves in computer game & 5 & 0.05 & 0.12 & 0.22 & 0.20 & 0.30 & 0.40 & 3\\
\hline
Explain moves in computer game & 10 & 0.30 & 0.35 & 0.40 & 0.40 & 0.45 & 0.50 & 3\\
\hline
\end{longtable}

\begin{longtable}{|p{4cm}|p{2cm}|p{1cm}|p{1cm}|p{1cm}|p{1.4cm}|p{1cm}|p{1cm}|r|}

\caption{Distribution of AI development milestones forecasts, 2016 survey, panel sample, fixed probabilities framing} \\

\hline
Milestones & Probability & Min year & Q1 year & Mean year & Median year & Q3 year & Max year & $n$\\
\hline
Assemble LEGO & 0.1 & 0 & 3.8 & 10.0 & 5.0 & 11.2 & 30 & 4\\
\hline
Assemble LEGO & 0.5 & 0 & 7.5 & 21.3 & 12.5 & 26.2 & 60 & 4\\
\hline
Assemble LEGO & 0.9 & 2 & 11.8 & 36.8 & 22.5 & 47.5 & 100 & 4\\
\hline
Compose Top-40 song & 0.1 & 3 & 4.5 & 7.0 & 7.5 & 10.0 & 10 & 4\\
\hline
Compose Top-40 song & 0.5 & 6 & 9.0 & 17.8 & 15.0 & 23.8 & 35 & 4\\
\hline
Compose Top-40 song & 0.9 & 12 & 14.2 & 31.8 & 32.5 & 50.0 & 50 & 4\\
\hline
\end{longtable}

\begin{longtable}{|p{4cm}|p{2cm}|p{1cm}|p{1cm}|p{1cm}|p{1.4cm}|p{1cm}|p{1cm}|r|}

\caption{Distribution of AI development milestones forecasts, 2016 survey, panel sample, fixed years framing} \\
\hline
Milestones & Year from survey year & Min prob & Q1 prob & Mean prob & Median prob & Q3 prob & Max prob & $n$\\
\hline
Compose Top-40 song & 10 & 0.20 & 0.20 & 0.20 & 0.20 & 0.20 & 0.20 & 1\\
\hline
Compose Top-40 song & 20 & 0.30 & 0.30 & 0.30 & 0.30 & 0.30 & 0.30 & 1\\
\hline
Compose Top-40 song & 50 & 0.90 & 0.90 & 0.90 & 0.90 & 0.90 & 0.90 & 1\\
\hline
Explain moves in computer game & 10 & 0.00 & 0.15 & 0.20 & 0.30 & 0.30 & 0.30 & 3\\
\hline
Explain moves in computer game & 20 & 0.00 & 0.20 & 0.37 & 0.40 & 0.55 & 0.70 & 3\\
\hline
Explain moves in computer game & 50 & 0.00 & 0.25 & 0.50 & 0.50 & 0.75 & 1.00 & 3\\
\hline
\end{longtable}

\begin{longtable}[H]{|p{3.5cm}|p{1.5cm}|p{1.5cm}|p{1.5cm}|p{1.5cm}|p{1.5cm}|p{1.5cm}|}

\caption{Cross-sectional sample: year of reaching milestone with 25\%, 50\%, and 90\% probability, median parameters aggregation method} \\

\hline
Milestones & 25\% - 2016 & 25\% - 2019 & 50\% - 2016 & 50\% - 2019 & 75\% - 2016 & 75\% - 2019\\
\hline
Win World Series of Poker & 2019.6 &  & 2021.4 &  & 2023.8 & \\
\hline
Win at Angry Birds & 2019.1 & 2021.8 & 2020.6 & 2022.8 & 2022.7 & 2024.1\\
\hline
Win at StarCraft & 2020.5 &  & 2023.3 &  & 2027.2 & \\
\hline
Phone banking services & 2020.3 & 2023.7 & 2022.4 & 2025.9 & 2025.2 & 2028.8\\
\hline
Find software vulnerability &  & 2022.6 &  & 2024.8 &  & 2028.1\\
\hline
Human-level translation & 2021.4 & 2023.6 & 2024.4 & 2025.5 & 2028.5 & 2028.1\\
\hline
\end{longtable}

\begin{longtable}[H]{|p{3.5cm}|p{1.25cm}|p{1.5cm}|p{1.5cm}|p{1.5cm}|p{1.5cm}|p{1.5cm}|}

\caption{Cross-sectional sample: year of reaching milestone with 25\%, 50\%, and 90\% probability, mean probability aggregation method} \\

\hline
Milestones & 25\% - 2016 & 25\% - 2019 & 50\% - 2016 & 50\% - 2019 & 75\% - 2016 & 75\% - 2019\\
\hline
Win World Series of Poker & 2017.9 &  & 2020.1 &  & 2025.0 & \\
\hline
Win at Angry Birds & 2017.3 & 2020.6 & 2019.4 & 2021.7 & 2023.5 & 2024.2\\
\hline
Win at StarCraft & 2018.7 &  & 2022.5 &  & 2033.3 & \\
\hline
Phone banking services & 2019.8 & 2022.8 & 2024.7 & 2026.2 & 2033.4 & 2031.6\\
\hline
Find software vulnerability &  & 2021.8 &  & 2025.0 &  & 2033.4\\
\hline
Human-level translation & 2019.9 & 2023.2 & 2024.4 & 2025.8 & 2034.8 & 2030.4\\
\hline
\end{longtable}

\begin{longtable}[H]{|p{3.5cm}|p{1cm}|p{1.5cm}|p{1.5cm}|p{1.5cm}|p{1.5cm}|p{1.5cm}|p{1.5cm}|}

\caption{Panel sample: year of reaching milestone with 25\%, 50\%, and 90\% probability, median parameters aggregation method} \\

\hline
Milestones & $n$ & 25\% - 2016 & 25\% - 2019 & 50\% - 2016 & 50\% - 2019 & 75\% - 2016 & 75\% - 2019 \\
\hline
Assemble LEGO & 4 & 2029.9 & 2026.1 & 2035.2 & 2029.3 & 2041.8 & 2033.5\\
\hline
Compose Top-40 song & 5 & 2023.6 & 2028.9 & 2026.9 & 2032.3 & 2031.3 & 2036.4\\
\hline
Explain moves in computer game & 5 & 2041.0 & 2032.5 & 2055.9 & 2039.7 & 2076.1 & 2049.2\\
\hline
Fold laundry & 5 & 2021.4 & 2027.1 & 2024.6 & 2029.5 & 2029.0 & 2032.4\\
\hline
Group unseen objects & 2 & 2030.5 & 2241.8 & 2037.9 & 2404.6 & 2047.4 & 2633.9\\
\hline
Human-level translation & 10 & 2020.8 & 2026.4 & 2023.3 & 2029.6 & 2026.5 & 2033.7\\
\hline
One-shot learning & 6 & 2022.5 & 2026.7 & 2026.6 & 2030.1 & 2032.1 & 2034.5\\
\hline
Output virtual world equations & 5 & 2029.9 & 2026.3 & 2046.0 & 2033.9 & 2072.0 & 2046.3\\
\hline
Perform well in Putnam Competition & 2 & 2127.1 & 2061.0 & 2176.4 & 2092.2 & 2238.7 & 2136.5\\
\hline
Phone banking services & 3 & 2025.8 & 2047.8 & 2028.9 & 2059.1 & 2032.6 & 2073.1\\
\hline
Prove math theorems & 2 & 2057.7 & 2036.8 & 2076.0 & 2041.4 & 2099.0 & 2046.8\\
\hline
Text to speech voice actor & 8 & 2023.3 & 2026.4 & 2026.6 & 2029.6 & 2030.8 & 2033.7\\
\hline
Transcribe human speech & 6 & 2024.8 & 2024.6 & 2028.0 & 2027.4 & 2032.0 & 2031.1\\
\hline
Win at Angry Birds & 8 & 2021.8 & 2021.2 & 2024.9 & 2021.9 & 2029.2 & 2022.7\\
\hline
Win at Atari & 3 & 2018.3 & 2021.2 & 2019.3 & 2021.8 & 2020.5 & 2022.5\\
\hline
Write history essay & 7 & 2021.8 & 2028.5 & 2025.1 & 2033.8 & 2029.5 & 2040.9\\
\hline
Write NYT bestseller & 1 & 2048.3 & 2062.4 & 2066.3 & 2086.0 & 2090.1 & 2117.2\\
\hline
Write Python code & 6 & 2023.9 & 2025.2 & 2029.3 & 2028.5 & 2036.9 & 2032.9\\
\hline
\end{longtable}

\begin{longtable}[H]{|p{3.5cm}|p{1cm}|p{1.5cm}|p{1.5cm}|p{1.5cm}|p{1.5cm}|p{1.5cm}|p{1.5cm}|}

\caption{Panel sample: year of reaching milestone with 25\%, 50\%, and 90\% probability, mean probabilities aggregation method} \\

\hline
Milestones & $n$ & 25\% - 2016 & 25\% - 2019 & 50\% - 2016 & 50\% - 2019 & 75\% - 2016 & 75\% - 2019 \\
\hline
Assemble LEGO & 4 & 2018.9 & 2024.3 & 2027.6 & 2027.8 & 2047.4 & 2035.4\\
\hline
Compose Top-40 song & 5 & 2024.6 & 2028.0 & 2032.4 & 2032.7 & 2048.0 & 2039.3\\
\hline
Explain moves in computer game & 5 & 2033.7 & 2025.0 & 2052.4 & 2032.4 & 2234.8 & 2047.0\\
\hline
Fold laundry & 5 & 2019.0 & 2023.0 & 2021.8 & 2026.8 & 2026.3 & 2035.6\\
\hline
Group unseen objects & 2 & 2021.0 & 2027.0 & 2032.6 & 2033.3 & 2066.3 & 2081.2\\
\hline
Human-level translation & 10 & 2020.5 & 2023.4 & 2024.0 & 2029.0 & 2033.0 & 2037.8\\
\hline
One-shot learning & 6 & 2020.6 & 2023.0 & 2025.7 & 2028.0 & 2038.8 & 2033.8\\
\hline
Output virtual world equations & 5 & 2017.7 & 2025.2 & 2021.8 & 2044.2 & 2134.3 & 2068.3\\
\hline
Perform well in Putnam Competition & 2 & 2091.3 & 2024.3 & 2155.6 & 2029.8 & 2226.5 & 2082.3\\
\hline
Phone banking services & 3 & 2028.6 & 2030.8 & 2042.7 & 2046.3 & 2056.5 & 2072.4\\
\hline
Prove math theorems & 2 & 2025.6 & 2023.2 & 2031.7 & 2029.0 & 2063.2 & 2059.7\\
\hline
Text to speech voice actor & 8 & 2024.1 & 2022.8 & 2029.4 & 2026.0 & 2037.7 & 2030.3\\
\hline
Transcribe human speech & 6 & 2023.0 & 2023.2 & 2027.3 & 2027.2 & 2034.6 & 2034.3\\
\hline
Win at Angry Birds & 8 & 2018.7 & 2020.3 & 2022.2 & 2021.2 & 2027.1 & 2022.8\\
\hline
Win at Atari & 3 & 2017.9 & 2020.8 & 2018.8 & 2021.6 & 2020.2 & 2023.0\\
\hline
Write history essay & 7 & 2020.6 & 2027.2 & 2026.2 & 2034.8 & 2037.0 & 2055.2\\
\hline
Write NYT bestseller & 1 & 2048.3 & 2062.3 & 2066.2 & 2085.9 & 2090.0 & 2117.2\\
\hline
Write Python code & 6 & 2019.0 & 2022.8 & 2024.6 & 2026.8 & 2030.8 & 2032.7\\
\hline
\end{longtable}

\begin{table}[H]
\caption{Randomization inference test results for testing difference between 2016 and 2019 AI progress milestone forecasts, Cross-Sectional Sample. The test statistic used is the Wasserstein metric that measures the absolute difference between the two forecast aggregate CDFs (generated using the median parameters method) for the interval between the years 2019 and 2119. $p$-values less than 0.05 after adjust the $p$-values for multiple comparisons using the Holm method are marked with a $^*$.\label{tab:cross-ri-ws}}

\begin{tabular}{|p{4cm}|p{3cm}|p{3cm}|p{2cm}|p{2cm}|}
\hline
Milestone & Wasserstein metric observed value & Wasserstein metric $p$-value & \textit{n} 2016 survey & \textit{n} 2019 survey\\
\hline
Assemble LEGO & 8.5661 & 0.1024 & 35 & 34\\
\hline
Compose Top-40 song & 8.1493 & 0.5946 & 38 & 25\\
\hline
Explain moves in computer game & 4.8521 & 0.5060 & 38 & 37\\
\hline
Fold laundry & 1.9045 & 0.8072 & 30 & 36\\
\hline
Group unseen objects & 12.4268 & 0.0002 & 29 & 41\\
\hline
Human-level translation & 1.7318 & 0.8992 & 42 & 32\\
\hline
One-shot learning & 1.9683 & 0.8464 & 32 & 40\\
\hline
Output virtual world equations & 9.5724 & 0.3152 & 51 & 30\\
\hline
Perform well in Putnam Competition & 26.6648 & 0.0120 & 45 & 45\\
\hline
Phone banking services & 3.3565 & 0.4458 & 31 & 43\\
\hline
Prove math theorems & 8.6718 & 0.2684 & 31 & 40\\
\hline
Text to speech voice actor & 2.1030 & 0.6862 & 43 & 38\\
\hline
Transcribe human speech & 2.3096 & 0.6162 & 33 & 30\\
\hline
Win at Angry Birds & 1.6160 & 0.8926 & 39 & 40\\
\hline
Win at Atari & 11.1468 & 0.0012 & 37 & 36\\
\hline
Write history essay & 1.4574 & 0.8914 & 42 & 36\\
\hline
Write NYT bestseller & 22.1726 & 0.0550 & 27 & 35\\
\hline
Write Python code & 4.4917 & 0.4686 & 36 & 34\\
\hline
\end{tabular}
\end{table}

\begin{table}[H]
\caption{Randomization inference test results for testing difference between 2016 and 2019 AI progress milestone forecasts, Cross-Sectional Sample. The test statistic used is the Kolmogorov–Smirnov statistic (2-sided) that quantifies the maximum absolute difference between the two forecast aggregate CDFs (generated using the median parameters method) in the interval between the years 2019 and 2119. $p$-values less than 0.05 after adjust the $p$-values for multiple comparisons using the Holm method are marked with a $^*$.\label{tab:cross-ri-ks-2}}

\begin{tabular}{|p{4cm}|p{3cm}|p{3cm}|p{2cm}|p{2cm}|}
\hline
Milestone & K-S test statistics (2-sided) observed value & K-S test statistics (2-sided) $p$-value & n 2016 survey & n 2019 survey\\
\hline
Assemble LEGO & 0.3451 & 0.3002 & 35 & 34\\
\hline
Compose Top-40 song & 0.3127 & 0.5944 & 38 & 25\\
\hline
Explain moves in computer game & 0.2212 & 0.5774 & 38 & 37\\
\hline
Fold laundry & 0.2793 & 0.6624 & 30 & 36\\
\hline
Group unseen objects & 0.5102 & 0.1968 & 29 & 41\\
\hline
Human-level translation & 0.0617 & 0.9378 & 42 & 32\\
\hline
One-shot learning & 0.0762 & 0.8578 & 32 & 40\\
\hline
Output virtual world equations & 0.2756 & 0.3622 & 51 & 30\\
\hline
Perform well in Putnam Competition & 0.6628 & 0.0102 & 45 & 45\\
\hline
Phone banking services & 0.3898 & 0.4076 & 31 & 43\\
\hline
Prove math theorems & 0.2088 & 0.2658 & 31 & 40\\
\hline
Text to speech voice actor & 0.1779 & 0.8080 & 43 & 38\\
\hline
Transcribe human speech & 0.1944 & 0.7456 & 33 & 30\\
\hline
Win at Angry Birds & 0.4762 & 0.8706 & 39 & 40\\
\hline
Win at Atari & 0.6019 & 0.0814 & 37 & 36\\
\hline
Write history essay & 0.0105 & 0.9782 & 42 & 36\\
\hline
Write NYT bestseller & 0.5505 & 0.0736 & 27 & 35\\
\hline
Write Python code & 0.1923 & 0.5892 & 36 & 34\\
\hline
\end{tabular}
\end{table}

\begin{table}[H]
\caption{Randomization inference test results for testing difference between 2016 and 2019 AI progress milestone forecasts, Cross-Sectional Sample. The test statistic used is the Kolmogorov–Smirnov statistic (1-sided) that quantifies the maximum difference between the two forecast aggregate CDFs (generated using the median parameters method; 2016 aggregate CDF minus 2019 aggregate CDF) in the interval between the years 2019 and 2119. $p$-values less than 0.05 after adjust the $p$-values for multiple comparisons using the Holm method are marked with a $^*$.\label{tab:cross-ri-ks-1}}

\begin{tabular}{|p{4cm}|p{3cm}|p{3cm}|p{2cm}|p{2cm}|}
\hline
Milestone & K-S test statistics (1-sided) observed value & K-S test statistics (1-sided) p-value & n 2016 survey & n 2019 survey\\
\hline
Assemble LEGO & -0.0122 & 0.9992 & 35 & 34\\
\hline
Compose Top-40 song & -0.0102 & 0.7466 & 38 & 25\\
\hline
Explain moves in computer game & -0.0026 & 0.9868 & 38 & 37\\
\hline
Fold laundry & 0.0000 & 0.7648 & 30 & 36\\
\hline
Group unseen objects & -0.0249 & 0.9998 & 29 & 41\\
\hline
Human-level translation & 0.0000 & 0.9846 & 42 & 32\\
\hline
One-shot learning & -0.0009 & 0.7752 & 32 & 40\\
\hline
Output virtual world equations & -0.0792 & 0.9866 & 51 & 30\\
\hline
Perform well in Putnam Competition & -0.2715 & 0.9982 & 45 & 45\\
\hline
Phone banking services & 0.3898 & 0.4188 & 31 & 43\\
\hline
Prove math theorems & 0.0149 & 0.8222 & 31 & 40\\
\hline
Text to speech voice actor & 0.0000 & 0.9770 & 43 & 38\\
\hline
Transcribe human speech & 0.0000 & 0.9814 & 33 & 30\\
\hline
Win at Angry Birds & 0.0000 & 0.9530 & 39 & 40\\
\hline
Win at Atari & -0.0040 & 1.0000 & 37 & 36\\
\hline
Write history essay & 0.0000 & 0.9184 & 42 & 36\\
\hline
Write NYT bestseller & -0.2406 & 0.9882 & 27 & 35\\
\hline
Write Python code & -0.0036 & 0.9358 & 36 & 34\\
\hline
\end{tabular}
\end{table}

\begin{table}[H]
\caption{Robust tests for the equality of variances for the 2016 versus 2019 milestone forecasts, cross-sectional sample, fixed-probabilities . We used the regression method described in \citep{iachine2010robust} for this analysis. For Model 1, the outcome variable is the absolute difference between each predicted year and the median predicted year for the milestone and survey year. For Model 2, the outcome variable is the absolute difference between each predicted year and the mean predicted year for the milestone and survey year. We clustered the standard errors by respondent-milestone.\label{tab:robust-cross-fp}}
\centering
\begin{tabular}[t]{lcc}
\toprule
  & Model 1: median & Model 2: mean\\
\midrule
(Intercept) & 14.254*** & 16.593***\\
 & (2.692) & (2.510)\\
Survey year: 2019 & 66.177 & 124.513*\\
 & (61.149) & (59.221)\\
\midrule
Number of responses & 1945 & 1945\\
Number of clusters (respondent-milestone) & 649 & 649\\
\bottomrule
\multicolumn{3}{l}{\textsuperscript{} * p < 0.05, ** p < 0.01, *** p < 0.001}\\
\end{tabular}
\end{table}

\begin{table}[H]
\caption{Non‐parametric combination test global $p$-value results for testing difference between 2016 and 2019 AI progress milestone forecasts, Cross-Sectional Sample. The test statistic used is the Wasserstein metric that measures the absolute difference between the two forecast aggregate CDFs (generated using the median parameters method) for the interval between the years 2019 and 2119. For robustness, we combined the partial $p$-values using three different combining functions: Fisher's product function, the minimum combining function, and Liptak's normal combining function. \label{tab:npc-results}}
\begin{tabular}{|l|r|r|r|}
\hline
Test statistic & Fisher & Minimum & Liptak\\
\hline
Wasserstein metric & 0.0028 & 0.0034 & 0.066\\
\hline
Kolmogorov-Smirnov test statistics (2-sided) & 0.4122 & 0.1670 & 0.552\\
\hline
Kolmogorov-Smirnov test statistics (1-sided) & 1.0000 & 1.0000 & 1.000\\
\hline
\end{tabular}
\end{table}

\begin{table}[H]

\caption{Comparing panel 2016 respondents' milestone forecasts with panel 2019 respondents' milestone forecasts using randomization inference. \label{tab:panel-ri}}
\centering
\begin{tabular}[t]{|p{5cm}|r|r|l|}
\hline
Outcome & Results & \textit{p}-value & Holm correction (\textit{p} < 0.05)\\
\hline
Wasserstein metric observed value & 2.558 &  & \\
\hline
Wasserstein metric, $p$-value (1-tailed) & 0.682 & 0.682 & No\\
\hline
Wasserstein metric, $p$-value (2-tailed) & 0.682 & 0.682 & No\\
\hline
Kolmogorov-Smirnov statistic, 2-sided, observed value & 0.239 &  & \\
\hline
Kolmogorov-Smirnov statistic, 2-sided, $p$-value (1-tailed) & 0.634 & 0.634 & No\\
\hline
Kolmogorov-Smirnov statistic, 2-sided, $p$-value (2-tailed) & 0.634 & 0.634 & No\\
\hline
Kolmogorov-Smirnov statistic, 1-sided, observed value & 0.000 &  & \\
\hline
Kolmogorov-Smirnov statistic, 1-sided, $p$-value (1-tailed) & 0.935 & 0.935 & No\\
\hline
Kolmogorov-Smirnov statistic, 1-sided, $p$-value (2-tailed) & 0.803 & 0.803 & No\\
\hline
\end{tabular}
\end{table}

\begin{table}[H]
\caption{Predicting log(year when milestone is at 50\%) using log(year when probability of HLMI is at 50\%). Model 1 uses fixed effects to account for variations between milestones. Model 2 uses random effects instead. The standard errors are clustered by respondent.}
\label{tab:hlmi-milestone-regression}

\begin{tabular}{p{4cm}p{3cm}lp{3cm}l}
\toprule                                              & Model 1: fixed effects &         & Model 2: random effects &         \\
Variable                                      & Coefficient (SE)       & $p$-value & Coefficient (SE)        & $p$-value \\
\midrule
Intercept                                     & 1.875 (0.701)          & 0.011   & 3.281 (1.654)           & 0.048   \\
Log(year when probability of HLMI is at 50\%) & 0.067 (0.165)          & 0.695   & 0.045 (0.164)           & 0.785   \\
\midrule
Number of forecasts                           & 782                    &         & 782                     &         \\
Number of respondents                         & 266                    &         & 266                     &   \\
\bottomrule
\end{tabular}
\end{table}

\end{document}